\documentclass[useAMS,usenatbib]{mn2e}
\usepackage{rotating}
\usepackage{times}
\usepackage{graphicx}
\usepackage{lscape}
\newcommand{\xmmn}{{\it XMM-Newton~\/}}
\newcommand{\asca}{{\it ASCA~\/}}
\newcommand{\chandra}{{\it Chandra~\/}}

\newcommand{\rosat}{{\it ROSAT~\/}}
\newcommand{\exosat}{{\it EXOSAT~\/}}

\newcommand{\einstein}{{\it Einstein~\/}}

\def\Msun{\hbox{$M_{\odot}$~\/}}

\def\ergsec{{\rm ~erg~s^{-1}}}
\def\chisq{{$\chi^{2}$}}

\def\delchi{{$\Delta \chi$~\/}}
\def\atpcm{{\rm ~atoms~cm^{-2}}}
\def\ctsec{{\rm ~count~s^{-1}}}

\def\H0{{\rm ~km~s^{-1}~Mpc^{-1}}}

\def\xspnorm{{\rm ~photon~cm^{-2}~s^{-1}~keV^{-1}}}

\def\eg{{\it e.g.,~\/}}

\def\ie{{\it i.e.,~\/}}

\def\la{\mathrel{\hbox{\rlap{\hbox{\lower4pt\hbox{$\sim$}}}{\raise2pt\hbox{$<$}}
}}}
\def\ga{\mathrel{\hbox{\rlap{\hbox{\lower4pt\hbox{$\sim$}}}{\raise2pt\hbox{$>$}}
}}}
\def\d25{D$_{25}$$~$}
\def\nh{{$N_{\rm H}$}$~$}

\def\hoii{Ho II X-1$~$}
\def\lx{L$_{\rm X}$$~$}

\def\arcmin{\hbox{$^\prime$~\/}}

\begin{document}

\title[\xmmn observations of the brightest ULXs] {\xmmn observations of the brightest Ultraluminous X-ray sources}
\author[A-M. Stobbart et al.]
    {A-M.\ Stobbart $^1$, T.P.\ Roberts $^1$ and J.\ Wilms $^2$
\\
$^1$X-ray \& Observational Astronomy Group, Dept. of Physics \&
Astronomy,
University of Leicester, Leicester LE1 7RH, U.K.\\
$^2$Department of Physics, University of Warwick, Coventry, CV4
7AL\\}

\date{Submitted}

\pagerange{\pageref{firstpage}--\pageref{lastpage}} \pubyear{2004}

\maketitle

\label{firstpage}

%=======================================================================

\begin{abstract}

We present an analysis of 13 of the best quality Ultraluminous X-ray
source (ULX) datasets available from \xmmn European Photon Imaging
Camera (EPIC) observations.  We utilise the high signal-to-noise in
these ULX spectra to investigate the best descriptions of their
spectral shape in the 0.3--10 keV range.  Simple models of an absorbed
power-law or multicolour disc blackbody prove inadequate at describing
the spectra.  Better fits are found using a combination of these two
components, with both variants of this model - a cool ($\sim 0.2$ keV)
disc blackbody plus hard power-law continuum, and a soft power-law
continuum, dominant at low energies, plus a warm ($\sim 1.7$ keV) disc
blackbody - providing good fits to 8/13 ULX spectra.  However, by
examining the data above 2 keV, we find evidence for curvature in the
majority of datasets (8/13 with at least marginal detections),
inconsistent with the dominance of a power-law in this regime.  In
fact, the most successful empirical description of the spectra proved
to be a combination of a cool ($\sim 0.2$ keV) classic blackbody
spectrum, plus a warm disc blackbody, that fits acceptably to 10/13
ULXs.  The best overall fits are provided by a physically
self-consistent accretion disc plus Comptonised corona model ({\sc
diskpn + eqpair}), which fits acceptably to 11/13 ULXs.  This model
provides a physical explanation for the spectral curvature, namely
that it originates in an optically-thick corona, though the accretion
disc photons seeding this corona still originate in an apparently cool
disc.  We note similarities between this fit and models of Galactic
black hole binaries at high accretion rates, most notably the model of
\citet{donek05}.  In this scenario the inner-disc and corona become
energetically-coupled at high accretion rates, resulting in a cooled
accretion disc and optically-thick corona.  We conclude that this
analysis of the best spectral data for ULXs shows it to be plausible
that the majority of the population are high accretion rate
stellar-mass (perhaps up to 80-$M_{\odot}$) black holes, though we
cannot categorically rule out the presence of larger, $\sim
1000$-\Msun intermediate-mass black holes (IMBHs) in individual
sources with the current X-ray data.

\end{abstract}


%=======================================================================

\section{Introduction}

\einstein X-ray observations were the first to reveal point-like,
extranuclear sources in some nearby galaxies with luminosities in
excess of $10^{39} \ergsec$ \citep{fabbiano89}. Subsequently, many of
these so-called Ultraluminous X-ray sources (ULXs) have displayed
short and long term variability, which suggests they are predominantly
accreting objects (see \citealt{milcol04} and references therein).
However, the observed luminosities of most ULXs exceed the Eddington
limit for spherical accretion onto a stellar-mass ($\sim$10-$
M_{\odot}$) black hole (BH). In fact, their luminosites are
intermediate between those of normal stellar mass BH X-ray Binaries
(BHBs) and Active Galactic Nuclei (AGN).  Therefore, the accretion of
matter onto {\it intermediate-mass\/} black holes (IMBHs, of
$\sim$$10^2$--$10^4$ $ M_{\odot}$) provide an attractive
explanation for ULXs, and could represent the long sought-after
`missing link' between stellar mass BHs and the supermassive BHs in
the nuclei of galaxies. However, the large populations of ULXs
associated with sites of active star formation (\eg in the Cartwheel
galaxy, \citealt{gao03}) demand rather too high formation rates of
IMBHs if they are to explain the ULX class as a whole \citep{king04}.
An alternative to accreting IMBHs is that ULXs may be a type of
stellar-mass BHB with geometrically \citep{king01} or relativistically
\citep*{koerding02} beamed emission, such that their intrinsic X-ray
luminosity does not exceed the Eddington limit.  Another possibility
is that they are stellar-mass BHBs that can achieve truly super
Eddington luminosities via slim \citep{ebisawa03} or radiation
pressure dominated \citep{begelman02} accretion discs.

As ULXs are probably the brightest class of X-ray binary fueled by the
accretion of matter onto a BH\footnote{Although the most likely
reservoir of fuel for an ULX is a companion star, others have been
suggested, for example the direct accretion of matter from molecular
clouds \citep{krolik04}.}, a knowledge of the properties of Galactic
BHBs could be vital in interpreting their characteristics.
Traditionally the X-ray spectra of BHBs have been fitted empirically
with two components, namely a power-law continuum and a multicolour
disc blackbody (MCD) component (\citealt{mitsuda84};
\citealt{makishima86}). In the standard picture, the power-law
component is thought to represent inverse-Compton scattering of
thermal photons from the accretion disc by hot electrons in a
surrounding corona. As such, the power-law represents the hard tail of
the X-ray emission while the MCD component models the soft X-ray
emission from the accretion disc.  The MCD model itself has been
formulated based on the best known model for accretion onto BHs (\ie
the thin accretion disc model, \citealt{shakura73}).

It has long been recognised that Galactic BHBs demonstrate various
X-ray spectral states which are defined by the balance of these two
components (\ie power-law and MCD) at any one time.  The three most
familiar X-ray bright states are the low/hard (LH), high/soft (HS,
also described as `thermal dominated') and the very high (VH, or
`steep power-law') states (see \citealt{mcclintock03} for further
details).  At lower mass accretion rates, a BHB usually enters the LH
state where their X-ray emission is dominated by a hard power-law
component ($\Gamma$$\sim$1.7), thought to arise from Comptonisation of
soft photons by a hot optically thin corona\footnote{However this is
still a topic of debate, with the main alternative for the X-ray
power-law emission being synchrotron emission from the radio jet that
is associated with this state (\eg \citealt{falcke95};
\citealt*{markoff01}).}.  In this state, the disc is either undetected
(\eg \citealt{belloni99}) or appears truncated at a much larger inner
radius and hence cooler than the parameters derived for the soft state
(\citealt{wilms99}, \citealt{mcclintock01}).  The soft X-ray state is
generally seen at a higher luminosity (\ie the HS state) and is best
explained as $\sim$1 keV thermal emission from a multi-temperature
accretion disc (\ie modelled with a MCD component). In this state, the
spectrum may also display a hard tail that contributes a small
percentage of the total flux.  The VH state is in many cases the most
luminous state and is characterised by an unbroken power-law spectrum
extending out to a few hundred keV or more.  The photon index is
typically steeper ($\geq 2.5$) than found in the LH state and
generally coincides with the onset of strong X-ray quasi-periodic
oscillations (QPOs).  A MCD component may also be present in the VH
state and the \exosat era demonstrated that some of the QPOs occur
when both disc and power-law components contribute substantial
luminosity \citep{vanderklis95}.

The idea of ULXs as analogues to Galactic BHBs in the HS state was
supported by \asca observations, which revealed that their 0.5--10 keV
spectra were successfully fitted with the MCD model with relatively
high disc temperatures (1.0--1.8 keV, \citealt{makishima00}).  As
such, the ULXs were considered to be mass-accreting BHs with the X-ray
emission originating in an optically-thick accretion disc.  In fact,
the use of the MCD model to describe these spectra permits one to
obtain an `X-ray--estimated' BH mass, $M_{\rm XR}$, from the
following equation (cf. \citealt{makishima00} equations (5)--(8)).

\begin{equation}
M_{\rm XR} = {{\xi\kappa^2}\over{8.86\alpha}} \frac{D}{\sqrt{{\rm cos}
i}} \sqrt{\frac{f_{\rm bol}}{2\sigma T^4_{\rm in}}} \; M_{\odot}
\end{equation}

Where $D$ is the distance to the X-ray source, which has an
inclination $i$, a full bolometric luminosity (from the MCD model) of
$f_{\rm bol}$ and an observed maximum disc colour temperature $T_{\rm
in}$.  In addition $\sigma$ is the Stefan-Boltzmann constant, $\kappa$
is the ratio of the colour temperature to the effective temperature
(`spectral hardening factor'), and $\xi$ is a correction factor
reflecting the fact that $T_{\rm in}$ occurs at a radius somewhat
larger than $R_{\rm in}$ (here, we assume that $R_{\rm in}$ is at the
last stable Keplerian orbit).  \cite{makishima00} use values of $\xi =
0.412$ and $\kappa = 1.7$, though other work has found different
values for the spectral hardening factor (e.g. $\kappa = 2.6$ for GRO
J1655-40, \citealt{st03}).  Finally, $\alpha$ is a positive parameter
with $\alpha = 1$ corresponding to a Schwarzschild BH.  However, the
masses inferred from the \asca data and Equation (1) are far too low
to be compatible with the large BH masses suggested by their
luminosities (assuming Eddington-limited accretion), for standard
accretion discs around Schwarzchild BHs.  \citet{makishima00}
suggested that this incompatibility could be explained if the BHs were
in the Kerr metric (\ie rapidly rotating objects), allowing smaller
inner disc radii and hence higher disc temperatures.

\chandra observations have provided some support for the
\citet{makishima00} results, with some ULX spectra being consistent
with the MCD model (\eg \citealt{roberts02}).  However, \chandra also
revealed that some ULX spectra showed a preference for a power-law
continuum rather than the MCD model (\eg \citealt{strickland01};
\citealt{roberts04}; \citealt{terashima04}). It has been suggested
that this preference for a power-law spectrum could be interpreted in
terms of the LH state seen in Galactic BHB candidates,
relativistically beamed jets or emission from a Comptonised accretion
disc in the VH state.  As well as these single component models, \asca
and \chandra spectroscopy have also suggested the presence of two
component spectra for some ULXs, comprising a MCD with a power-law
component.  For example, previous \asca analyses hinted at evidence
for IMBHs, \ie cool accretion disc components (see below), but these
observations were not sensitive enough to statistically require two
component modelling (\eg \citealt{colbert99}).  Similar results have
been obtained by {\it Chandra}, \eg ULXs in NGC 5408
\citep{kaaret03} and NGC 6946 \citep{roberts03}.  Conversely,
\chandra spectra of the Antennae ULXs (\citealt{zezas02a};
\citealt{zezas02b}) revealed an accretion disc (MCD) component
consistent with the high temperature \asca results (\ie $kT_{\rm
in}$$\sim$1 keV), together with a hard power-law component
($\Gamma$$\sim$1.2).

It is only recently, using high quality {\it XMM-Newton}/EPIC
spectroscopy of ULXs, that it has been demonstrated that the
addition of a soft thermal disc component to a power-law continuum
spectrum provides a strong statistical improvement to the best
fitting models to ULX data (\eg \citealt{miller03};
\citealt*{miller04a}). These particular observations have provided
strong support for the IMBH hypothesis by revealing disc temperatures
in these sources up to 10 times lower than commonly measured in
stellar mass BHBs, consistent with the expectation for the accretion
disc around a $\sim$1000-\Msun IMBH\footnote{It is common for the generic
range of masses for IMBHs to be quoted as 20--$10^6$
\Msun.  The lower limit comes from a consideration of the measured
masses of BHs in our own Galaxy \citep{mcclintock03}, and a
theoretical limit for the mass of a BH formed from a single massive
star \citep{fryer01}.  However, more recent population synthesis
analyses show that BHs of up to $\sim 80$-\Msun may be formed in young
stellar populations \citep{belczynski04}.  Hence, when we refer to
IMBHs in this paper we refer specifically to the larger
$\sim$1000-\Msun IMBHs implied by the cool accretion disc
measurements.} (cf. Equation (1)).  However, in a few cases, {\it
XMM-Newton}/EPIC observations have revealed a more unusual
two-component X-ray spectrum.  A detailed analysis of such a source is
presented in \citet*{stobbart04}. In this case, the ULX is the
brightest X-ray source in the nearby (1.78 Mpc) Magellanic-type galaxy
NGC 55 and, although its X-ray luminosity only marginally exceeds
$10^{39} \ergsec$, it represents one of the highest quality ULX
datasets obtained to date. The initial low flux state data were best
fitted with an absorbed power-law continuum ($\Gamma$$\sim$4), while a
subsequent flux increase was almost entirely due to an additional
contribution at energies $> 1$ keV, adequately modelled by a MCD
component ($kT_{\rm in}$$\sim$0.9 keV).

Whilst this accretion disc component is reasonable for stellar-mass
BHs, the dominance of the power-law continuum at soft X-ray energies
is problematic.  Such a soft power-law cannot represent Comptonised
emission from a hot corona, as one would not expect to see the coronal
component extend down below the peak emissivity of the accretion disc,
where there would be insufficient photons to seed the corona.
Alternative sources of seed photons for the corona are unlikely; for
example, the incident photon flux of the secondary star at the inner
regions of the accretion disc is too low to provide the seeding
(cf. \citealt{roberts05}).  It also seems unlikely that the power-law
emission could arise from processes at the base of a jet, as these are
typically represented by much harder photon indices than measured here
($\Gamma$$\sim$1.5--2; \citealt{mnw05} and references therein).
Indeed, with the possible exception of NGC 5408 X-1 \citep{kaaret03},
there is no evidence that ULXs do display bright radio jets, though
this cannot be excluded by current observations \citep{kcf05}.  The
possibility of the soft component resulting from an outflow of
material from the accretion disc may also be discounted as this would
produce a thermal spectrum rather than a power-law continuum.

Although this spectral description has not been seen in Galactic
systems, a second case has been reported independently for the nearest
persistent extragalactic ULX (M33 X-8) by \citet{foschini04}.  The
non-standard model provided the best fit to this ULX with
$\Gamma$$\sim$2.5 and $kT_{\rm in}$$\sim$1.2 keV.  However, this
source is also at the low luminosity end of the ULX regime with
\lx$\sim$2$\times 10^{39}\ergsec$. A further possible case, in a more
luminous ULX, has arisen from the \xmmn data analysis of NGC 5204 X-1
\citep{roberts05}.  In this case the authors show that there is
spectral ambiguity between the non-standard fit ($\Gamma$$\sim$3.3,
$kT_{\rm in}$$\sim$ 2.2 keV) and the IMBH model ($\Gamma$$\sim$2.0,
$kT_{\rm in}$$\sim$ 0.2 keV), with both providing statistically
acceptable fits to the data.  Even more recently, two additional
examples of this spectral form have been uncovered in an \xmmn survey
of ULXs by \citet{feng05}.

Although it is difficult to derive a literal physical interpretation
from the non-standard model, it does provide an accurate empirical
description of ULX spectra in some cases, and as such it has the
potential to provide new insights into the nature of these
sources. Therefore in this paper we re-evaluate current data in an
attempt to determine the best spectral description for the shape of
high quality ULX spectra, and ask what consequences this has for the
idea of ULXs as accreting IMBHs. The paper is structured as follows:
Sec.~\ref{sec_sample} -- introduction to the ULX sample;
Sec.~\ref{sec_obs} -- details of the observations and data reduction;
Sec.~\ref{sec_spectra} -- description of the spectral analysis;
Sec.~\ref{sec_lx_kt} -- a comment on the luminosity and inner disc
temperature relationship of these ULXs; Sec.~\ref{sec_discussion} -- a
discussion of our results; and finally Sec.~\ref{sec_conclusions} --
our conclusions.

%=======================================================================
\section{The Sample}
\label{sec_sample}

As our primary goal is to find the best description(s) of the shape of
ULX spectra, only the highest quality datasets were chosen.  The ULXs
were initially selected from the \rosat catalogues of
\citet{roberts00} and \citet{colbert02} to provide a list of
historically-bright ULXs that are resolved at a spatial resolution
similar to {\it XMM-Newton}\footnote{The \rosat lists were deemed
appropriate as most bright ULXs are persistent and vary by factors of
no more than 2--3 in flux over a baseline of years,
cf. \citet{roberts04}.}.  We applied a source selection criteria of
observed count rates of $>$ 10 counts ks$^{-1}$ in the \rosat HRI
camera, combined with $>$ 10 ks of {\it XMM-Newton}/EPIC data
available in the archive by December 2004, to select ten ULXs with
potentially sufficient counts for very detailed spectral analysis.  In
addition, we included three more high quality ULX datasets:
proprietary data for Holmberg II X-1 (hereafter Ho II X-1), and two
sources not quite bright enough in the \rosat bandpass to be
classified as ULXs, namely the ULX in NGC 55 and M33 X-8.  Whilst some
of the sources in this sample have been observed more than once, we
have only selected the longest individual exposure in each case to
provide the clearest single view of their spectra.  Our final sample
of 13 sources from 12 different galaxies is listed in
Table~\ref{sample}. The selected ULXs are located at distances of
between 800 kpc and 17.8 Mpc, possess \xmmn count rates between 0.1
and 8.9 $\ctsec$ and cover the full range of ULX luminosities
($\sim$$10^{39} \ergsec$-- few $\times 10^{40}\ergsec$).  Hereafter we
refer to the sources by their names as given in column (1) of Table
\ref{sample}.

%=======================================================================
\begin{table*}
\begin{center}
\caption{\label{sample}The sample}
\begin{tabular}{llcccccc}
\hline
                    	&                       	&               &                   	&\nh                	&\it{d}         &\lx  \\
\hspace{3mm}Source  	&\hspace{5mm}Alternate names    &R.A. (J2000)   &DEC (J2000)        	&(10$^{20}$cm$^{-2}$)   &(Mpc)          &(10$^{39} \ergsec$) \\
\hspace{5mm}(1)     	&\hspace{13mm}(2)               &(3)            &(4)                	&(5)                    &(6)            &(7)\\
\hline
NGC 55 ULX $^{1}$       &XMMU J001528.9$-$391319 $^a$   &00 15 28.9     &$-$39 13 19.1 $^a$     &1.74                   &1.78 $^b$      &1.3\\
                	&NGC 55 6 $^c$              	&   		&           		&           		&       	&\\
                	&Source 7 $^d$              	&   		&           		&           		&       	&\\

M33 X-8 $^{2}$         	&--                      	&01 33 50.9     &$+$30 39 37.2 $^e$     &5.69                   &0.70 $^f$      &1.0\\

NGC 1313 X-1 $^{3}$     &Source 6 $^g$              	&03 18 20.0     &$-$66 29 11.0 $^h$     &3.96               	&3.70 $^i$      &4.7\\
               		&IXO 7 $^j$                 	&       	&           		&           		&       	&\\

NGC 1313 X-2 $^{3}$     &Source 4 $^g$              	&03 18 22.3     &$-$66 36 03.8 $^{k}$   &3.94               	&3.70 $^i$      &1.7\\
                	&IXO 8 $^j$                 	&       	&           		&           		&       	&\\

NGC 2403 X-1 $^{4}$     &Source 21 $^{l}$           	&07 36 25.5     &$+$65 35 40.0 $^{l}$   &4.17                   &4.20 $^f$      &2.7\\

\hoii $^{5}$		&IXO 31 $^j$         		&08 19 29.0     &$+$70 42 19.3 $^{m}$   &3.41                   &4.50 $^i$      &17\\

M81 X-9 $^{6}$  	&Holmberg {\small IX} X-1 $^{n}$&09 57 53.2     &$+$69 03 48.3 $^{o}$   &4.25                   &3.55 $^p$      &12\\
                	&IXO 34 $^j$                	&       	&           		&           		&       	&\\
                	&NGC 3031 10 $^q$           	&       	&           		&           		&       	&\\
                	&H 44 $^r$                  	&       	&           		&           		&       	&\\

NGC 3628 X-1 $^{4}$     &IXO 39 $^{j}$              	&11 20 15.8     &$+$13 35 13.6 $^{s}$   &2.22                   &7.70 $^f$      &5.2\\

NGC 4395 X-1 $^{4}$     &NGC 4395 X2 $^t$           	&12 26 01.5     &$+$33 31 30.5 $^{u}$   &1.36                   &3.60 $^f$      &0.6\\
                    	&IXO 53 $^j$                	&               &                   	&                   	&               &\\

NGC 4559 X-1 $^{4}$     &X-7 $^{v}$                 	&12 35 51.7     &$+$27 56 04.1 $^{w}$   &0.82                   &9.70 $^f$      &9.1\\
                	&IXO 65 $^j$               	&       	&           		&           		&       	&\\

NGC 4861 ULX $^{1}$     &IXO 73 $^j$                	&12 59 01.9     &$+$34 51 13.5 $^{x}$   &1.21                   &17.80 $^i$   	&8.8\\
            		&X1 $^y$            		&       	&           		&           		&       	&\\

NGC 5204 X-1 $^{4}$     &IXO 77 $^j$                	&13 29 38.6     &$+$58 25 05.7 $^{z}$   &1.38                   &4.80 $^f$      &4.4\\
                	&HST 3 $^{z}$               	&       	&           		&           		&       	&\\
                	&U1 $^{aa}$             	&       	&           		&           		&      	 	&\\

M83 ULX  $^{1}$         &Source 13 $^{bb}$          	&13 37 19.8     &$-$29 53 48.9 $^{cc}$  &3.69                   &4.70 $^i$      &1.0\\
                	&IXO 82 $^j$                	&       	&           		&           		&       	&\\
                	&H30 $^{dd}$            	&       	&           		&           		&       	&\\
                	&H2 $^{ee}$                 	&       	&           		&           		&       	&\\
\hline
\end{tabular}
\begin{minipage}[t]{6.2in}
{\sc Notes}: 
(1) Source designation; 
(2) Alternative names; 
(3--4) X-ray source position from \xmmn and \chandra data, or position of
possible optical counterpart; 
(5) Galactic absorption column density
from the `{\sc nh}' {\sc ftools} program (based on the measurements of
\citealt{dickey90}); 
(6) Distance to the host galaxy; 
(7) Observed
X-ray luminosity (0.3--10 keV) based on the results of the physically
self-consistent modelling (see later). 
{\sc References}: 
$^{1}$ This paper, 
$^{2}$ \citet{markert83}, 
$^{3}$ \citet{colbert95}, 
$^{4}$ \citet{roberts00}, 
$^{5}$ \citet{dewangan04}, 
$^{6}$ \citet{fabbiano88}, 
$^a$ \citet{stobbart04}, 
$^b$ \citet{kara03}, 
$^c$ \citet*{read97}, 
$^d$ \citet*{schlegel97}, 
$^e$ \citet{foschini04}, 
$^f$ \citet*{ho97}, 
$^g$ \citet{schlegel00},
$^h$ \citet{miller03}, 
$^i$ \citet{tully88}, 
$^j$ \citet{colbert02}, 
$^k$ \citet{zampieri04}, 
$^l$ \citet{schlegel03}, 
$^m$ \citet*{kaaret04}, 
$^n$ \citet{miller04a}, 
$^o$ \citet{ramsey06}, 
$^p$ \citet{paturel02}, 
$^q$ \citet{radecke97}, 
$^r$ \citet{immler01},
$^s$ \citet{strickland01}, 
$^t$ \citet*{lira00}, 
$^u$ Sourcelist (Obs.Id 5302) via {\it Chandra X-ray Centre:}
http://cxc.harvard.edu/chaser, 
$^v$ \citet*{vogler97}, 
$^w$ \citet{cropper04}, 
$^x$ Sourcelist (Obs.Id 014115010) via {\it XMM-Newton Science Archive:}
http://xmm.vilspa.esa.es/external/xmm\_data\_acc/xsa/index.shtml,
$^y$ \citet{liu05}, 
$^z$ \citet{goad02}, 
$^{aa}$ \citet*{liu04}, 
$^{bb}$ \citet{ehle98}, 
$^{cc}$ \citet{soria02},
$^{dd}$ \citet{immler99}, 
$^{ee}$ \citet*{trinchieri85}.
\end{minipage}
\end{center}
\end{table*}
%=======================================================================

%=======================================================================
\begin{table*}
\small{
\begin{center}
\caption{\label{obs}\xmmn Observation Log}
\begin{tabular}{lcccccccccc}

\hline
\hspace{3mm}Source&Obs.ID   &Date       &Duration (s)   &Net exp. (s)   &Rate (ct s$^{-1}$) 	&MOS-1  &MOS-2  &pn     &Position   &Ref\\
\hspace{5mm}(1) &(2)        &(3)        &(4)            &(5)                    &(6)            &(7)    &(8)    &(9)    &(10)        &(11)\\
\hline
NGC 55 ULX      &0028740201 &2001-11-14 &34025          &30410          &2.08           	&FF &FF &FF             &on-axis    &1\\
M33 X-8         &0102640101 &2000-08-04 &18672          &6850           &8.55           	&SW &SW &FF	        &on-axis    &2,3,4\\
NGC 1313 X-1    &0106860101 &2000-10-17 &42769          &18490          &1.15           	&FF &FF &FF             &on-axis    &4,5,6\\
NGC 1313 X-2    &0106860101 &2000-10-17 &42769          &18490          &0.39           	&FF &FF &FF             &off-axis$^a$  &4,5,6\\
NGC 2403 X-1    &0164560901 &2004-09-12 &84600          &57063          &0.47           	&FF &FF &FF             &on-axis    &4\\
\hoii 		&0200470101 &2004-04-15 &111999 	&47260          &4.52           	&LW &LW &FF             &on-axis    &7\\
M81 X-9         &0112521101 &2002-04-16 &11935          &8440           &3.25           	&FF &FF &FF             &on-axis    &4,5\\
NGC 3628 X-1    &0110980101 &2000-11-27 &60745          &45260          &0.20               	&FF &FF &FF             &on-axis    &4,5\\
NGC 4395 X-1    &0142830101 &2003-11-30 &118900         &98900          &0.18               	&FF &FF &FF             &on-axis    &4,8\\
NGC 4559 X-1    &0152170501 &2003-05-27 &43290          &37660          &0.48               	&SW &SW &FF             &on-axis    &4,9\\
NGC 4861 ULX    &0141150101 &2003-06-14 &29600          &14590          &0.11           	&FF &FF &FF             &on-axis    &--\\
NGC 5204 X-1    &0142770101 &2003-01-06 &32999          &17048          &0.87               	&FF &FF &FF             &on-axis    &4,10\\
M83 ULX         &0110910201 &2003-01-27 &31722          &21130          &0.21               	&FF &FF &EFF            &off-axis$^b$   &4\\
\hline

\end{tabular}
\begin{minipage}[t]{6.7in}
{\sc Notes}: (1) Source designation; (2) Observation identifier; (3)
Observation date (yyyy-mm-dd); (4) Observation duration (NB. this does
include calibration observations in some cases); (5) Useful exposure
after correcting for flaring episodes and ensuring simultaneous
operation of the EPIC cameras; (6) Combined EPIC count rates derived
from the combined X-ray light curves (0.3--10 keV); (7--9) Observing
mode of each EPIC detector (SW : Small Window, LW : Large Window, FF :
Full Frame, EFF : Extended Full Frame); (10) Source position with
respect to the centre of the pn field of view: {\it a}--$\sim$7
\arcmin offset; {\it b}-- $\sim$6.5 \arcmin offset; (11)
References for previous analyses of these datasets (although not all
references contain a complete analysis of the ULX):
1--\citet{stobbart04}, 2--\citet{foschini04}, 3--\citet{pietsch04},
4--\citet{feng05}, 5--\citet{wang04}, 6--\citet{miller03},
7--\citet{goad05}, 8--\citet{vaughan05}, 9--\citet{cropper04},
10--\citet{roberts05}.
\end{minipage}
\end{center}
}
\end{table*}
%=======================================================================

\section{Observations and data analysis}
\label{sec_obs}

In this work we have utilised data from the EPIC cameras on board {\it
XMM-Newton} (\citealt{turner01}; \citealt{struder01}). The datasets
were obtained through the \xmmn public data archive (excluding
proprietary \hoii data) and details of the observations are shown in
Table \ref{obs}. The data were processed and reduced using the
standard tools of {\sc xmm-sas} software v.6.0.0.  In some cases the
observations were affected by soft proton flaring for which
preliminary cleaning was necessary. For these observations we
extracted full field X-ray (0.3--10 keV) light curves and screened for
flaring using Good Time Interval (GTI) files based on either a time or
count rate criterion.  The NGC 55 ULX and NGC 5204 X-1 observations
were not affected by flaring episodes.  The NGC 2403 X-1 observation
was only affected by flaring at the end of the exposure (the last
$\sim$20 ks), therefore we used a time selection to exclude this
flaring event. For the remaining sources we used a count rate cut-off
criterion to produce a GTI file.  The exact value of the cut-off was
allowed to vary from field-to-field, to provide the best compromise in
each case between excluding high background periods and facilitating
the longest available exposure on the ULX.  In practise the actual
cut-off values varied in the 6.5--17.5 $\ctsec$ (0.3--10 keV) range.
In all cases we used only those data recorded when the MOS and pn
cameras were in simultaneous operation (see Table \ref{obs} for the
net good exposure times).

For each ULX, events were extracted in a circular aperture centred on
the source position given in Table \ref{sample}. The background was
taken from a circular region, near to the source in the pn camera and
at the same distance from the readout node. The chosen background
regions were the same in all three detectors where observations were
taken using the full-frame observing mode. However, in three cases the
MOS cameras were operated in either the large-window or small-window
mode, so here we used a background region closest to the one used for
the pn extraction. The size of the source and background spectral
extraction regions are listed in Table~\ref{extraction}.

For the pn camera the spectra were extracted using event patterns
0--4, which allows `single' and `double' pixel events.  We also set
`FLAG=0' to exclude all events at the edge of the CCD and events from
bad pixels. For the MOS detectors we used event patterns 0--12 which
allows `single', `double', `triple' and `quadruple' pixel events.  We
employed a less conservative screening criterion for the MOS data by
using the flag expression {\sc \#xmmea\_em} to exclude hot pixels and
events outside of the field of view.  The {\sc sas} task {\it
especget} was used to produce source and background spectra for each
ULX, together with the appropriate Redistribution Matrix File (RMF)
and Ancilliary Response File (ARF). Spectral files were grouped to
require at least 20 counts bin$^{-1}$ before fitting, to ensure Gaussian
statistics.

Model spectra were fitted to the data using the HEAsoft X-ray spectral
fitting package {\sc xspec} (v.11.3.0).  The pn, MOS-1 and MOS-2
spectra for each object were fitted simultaneously, but we included
constant multiplicative factors in each model to allow for calibration
differences between the cameras.  This value was frozen at unity for
the pn data and allowed to vary for the MOS detectors, with the values
typically agreeing within 20 per cent.  Quoted fluxes are an average
of the three measurements.

Spectra were initially fitted in the 0.3--10 keV band.  However, in
several cases we found that the data below 0.5 keV did not agree with
any of the models we tried, leaving large residuals (particularly with
respect to the pn data) to the best fits.  We note that the datasets
with this problem -- NGC 55 ULX, NGC 1313 X-1 \& X-2 and NGC 3628 X-1
-- were all obtained in the years 2000--2001 (in fact M33 X-8 is the
only observation from this epoch without these fitting residuals).  At
first glance this would appear to be a problem with the calibration at
early mission times; however the pn calibration at least has remained
remarkably static in orbit (EPIC team, priv. comm.).  As we cannot
explain this variation we conservatively excluded these data in all
EPIC cameras, such that spectral fitting was restricted to the 0.5--10
keV range in these four cases.  Also, in the case of \hoii
there is a possible pn calibration feature that is particularly
prominent at low energies \citep{goad05}, therefore we followed the
method of these authors and excluded the pn data below 0.7 keV to
account for this, whilst retaining MOS data down to 0.3 keV.

The X-ray spectra were modified for absorption following
\citet{balucinska92}, assuming the solar abundances of
\citet{anders89} (\ie the {\sc wabs} component in {\sc xspec}). We
used two absorption components, one of which was fixed for each source
to represent the appropriate foreground column density through our
Galaxy (\citealt{dickey90} -- see Table \ref{sample}), and the second
component was left free to fit the data to represent additional
absorption within the host galaxy and/or intrinsic to the ULX. The
errors quoted in this work are at the 90\% confidence level for one
interesting parameter.  Throughout this analysis we distinguish
statistically acceptable fits from unacceptable fits using a fixed
criterion of $P_{\rm rej} < 95$ per cent, where $P_{\rm rej}$ is the
probability of rejection derived directly from the $\chi^2$ statistic
for the spectral fit.

%=======================================================================
\begin{table}
\begin{center}
\caption{\label{extraction}Spectral extraction apertures}
\begin{tabular}{lcc}
\hline
            &\multicolumn{2}{c}{Extraction radius}\\
Source          &Source     &Background\\
(1)         &(2)        &(3)\\
\hline
NGC 55 ULX      &60     &75\\
M33 X-8         &35     &120\\
NGC 1313 X-1    &35     &70\\
NGC 1313 X-2    &35     &70\\
NGC 2403 X-1    &45     &45\\
\hoii    	&45     &60\\
M81 X-9         &30     &45\\
NGC 3628 X-1    &35     &70\\
NGC 4395 X-1    &40     &40\\
NGC 4559 X-1    &40     &80\\
NGC 4861 ULX    &15     &45\\
NGC 5204 X-1    &40     &50\\
M83 ULX         &35     &50\\
\hline
\end{tabular}
\begin{minipage}[t]{2in}
{\sc Notes}: (1) Source designation; (2)--(3) Aperture radii in
arcseconds
\end{minipage}
\end{center}
\end{table}

%=======================================================================
\section{Spectral properties}
\label{sec_spectra}

\subsection{Simple models}

The X-ray spectra appeared relatively smooth and featureless in each
case, hence we began by fitting simple continuum models to the
data. We tried the two models most often used in the past as
single-component descriptions of ULX spectra, namely a power-law
continuum model and the canonical MCD model. The details of these fits
are given in Table~\ref{single} and the quality of the X-ray spectra
is illustrated using the power-law fits in Figure~\ref{spec1}.  Due to
the excellent spectral quality, we were able to reject the power-law
model in 8 out of the 13 cases at $>$ 95\% confidence.  Interestingly,
the five ULXs for which a power-law is an adequate description of the
data (NGC 1313 X-1 \& X-2, M83 ULX, NGC 4861 ULX, NGC 5204 X-1)
include the three poorest quality datasets. Even more notably, the MCD
model does not provide an acceptable fit to any of the spectra.

%=======================================================================

\begin{figure*}
\begin{center}
\scalebox{0.35}{{\includegraphics[angle=270]{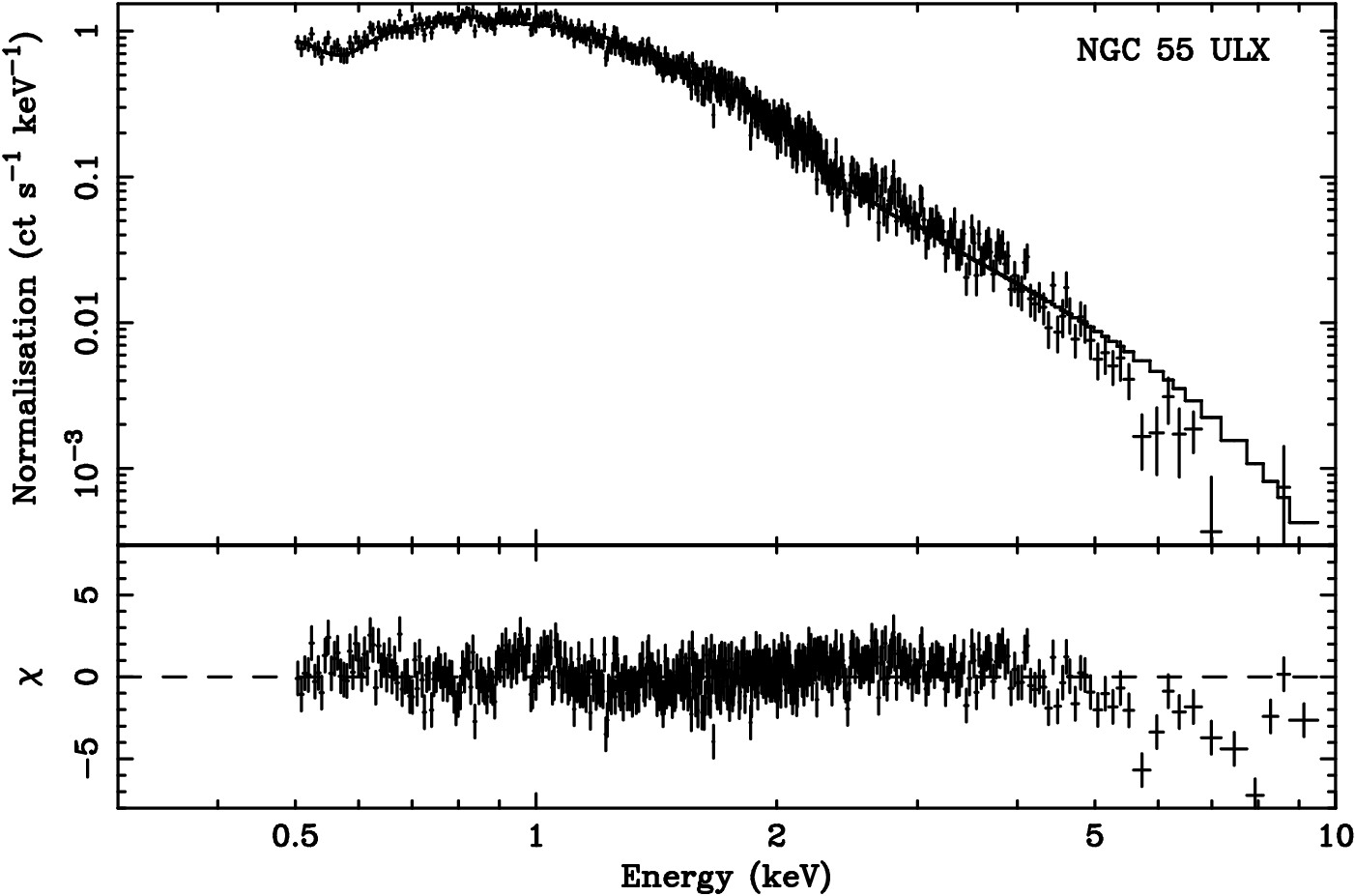}}}
\hspace*{0.5cm}
\scalebox{0.35}{{\includegraphics[angle=270]{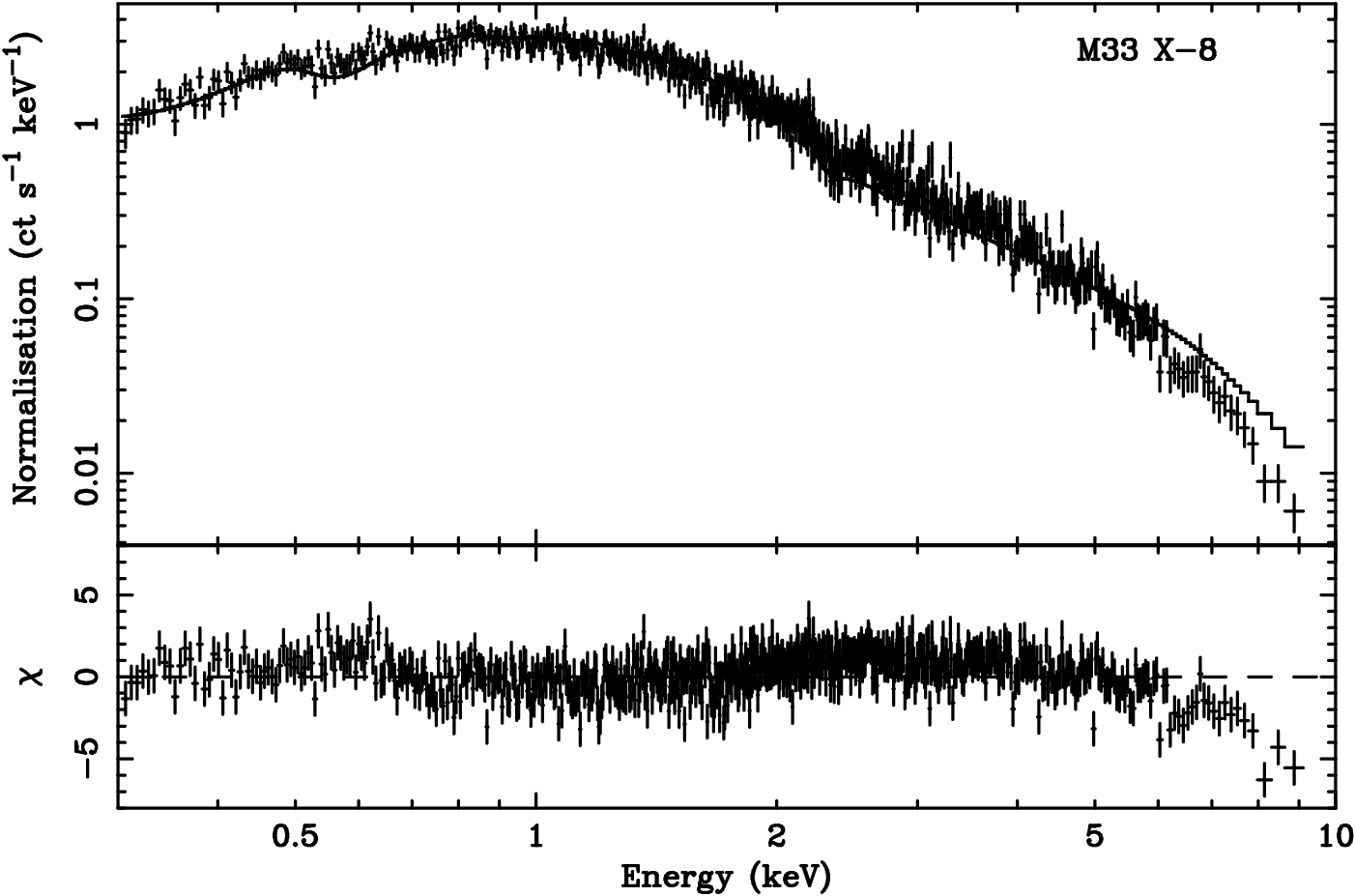}}}\vspace*{0.2cm}
\scalebox{0.35}{{\includegraphics[angle=270]{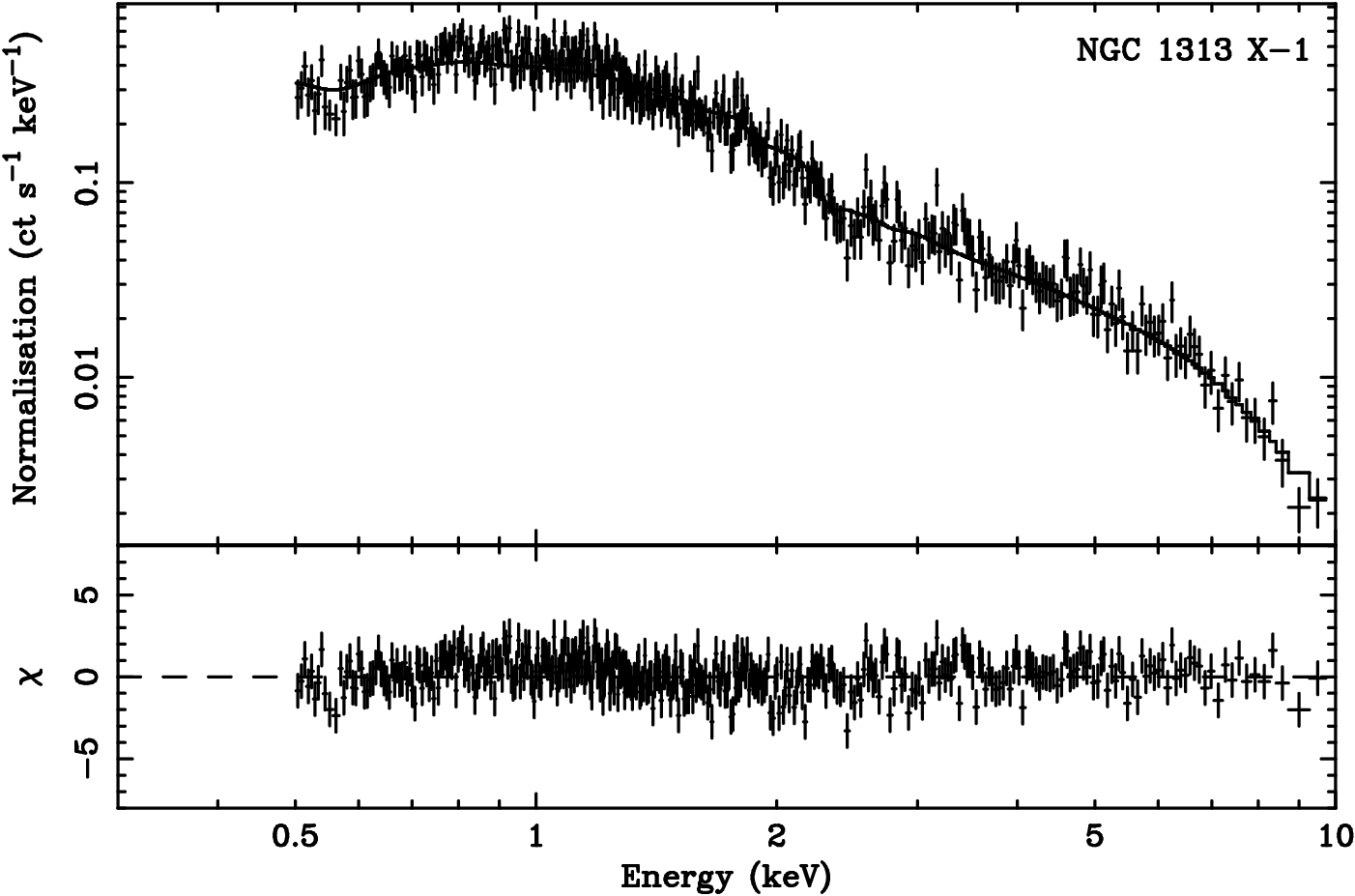}}}
\hspace*{0.5cm}
\scalebox{0.35}{{\includegraphics[angle=270]{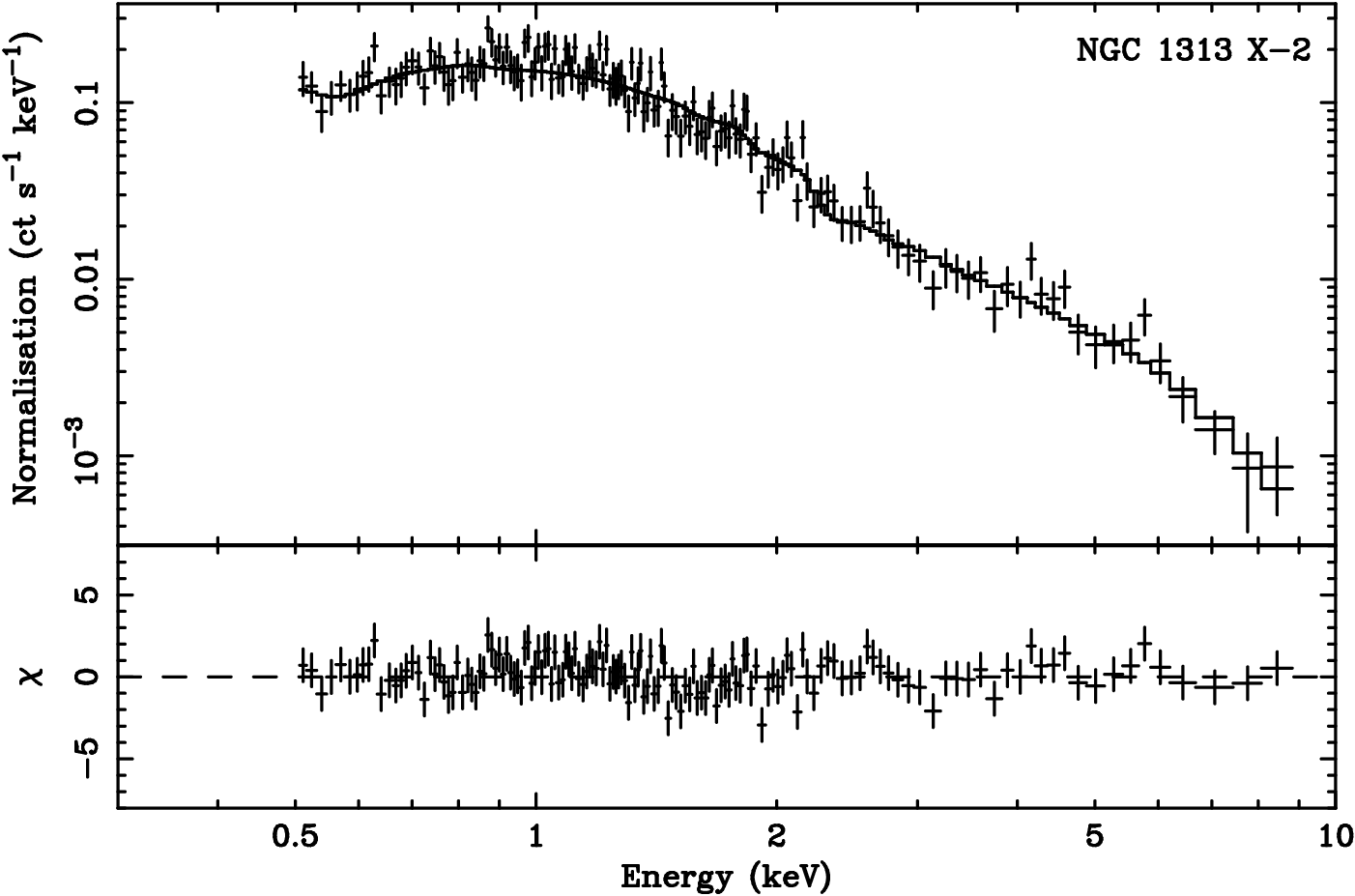}}}\vspace*{0.2cm}
\scalebox{0.35}{{\includegraphics[angle=270]{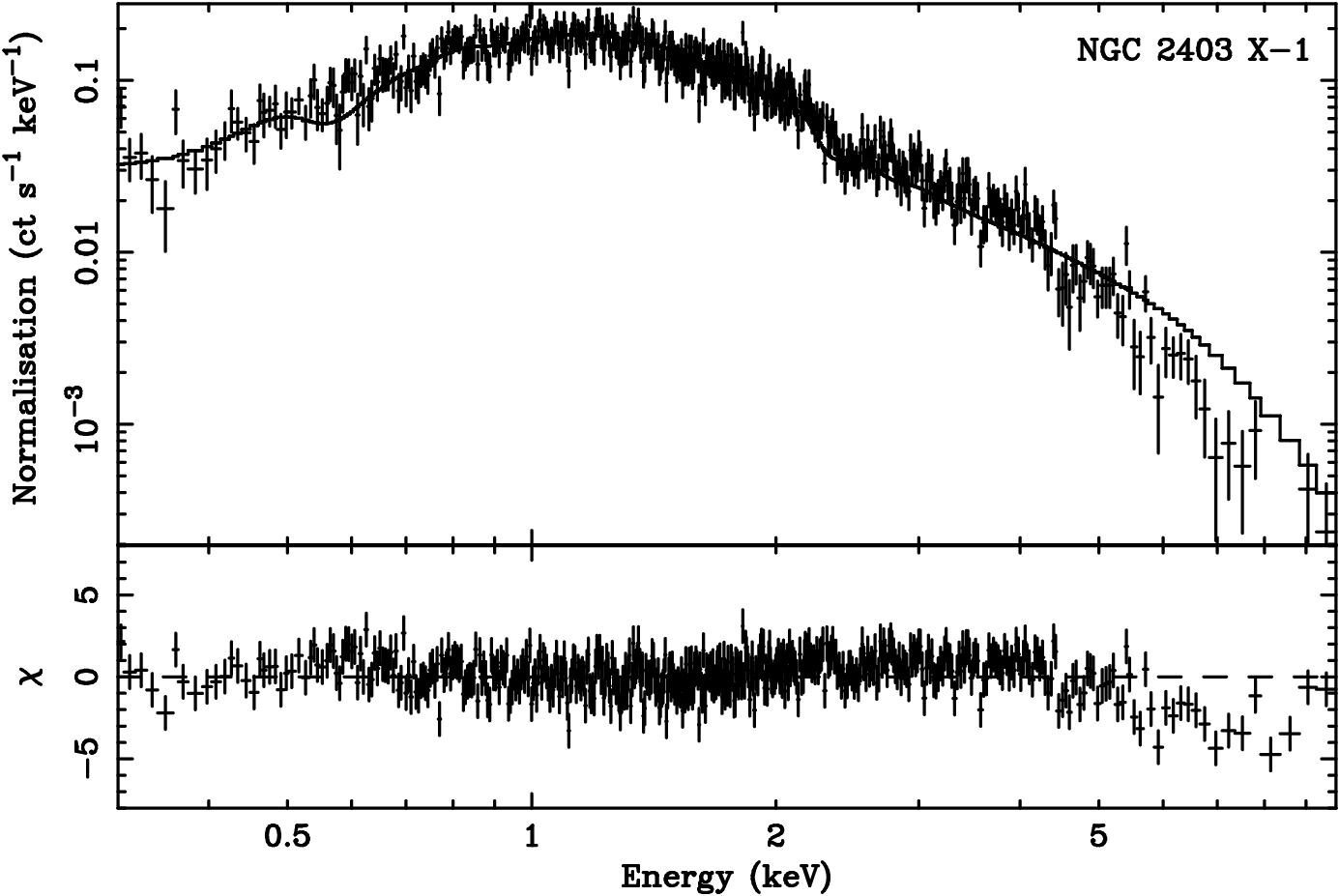}}}
\hspace*{0.5cm}
\scalebox{0.35}{{\includegraphics[angle=270]{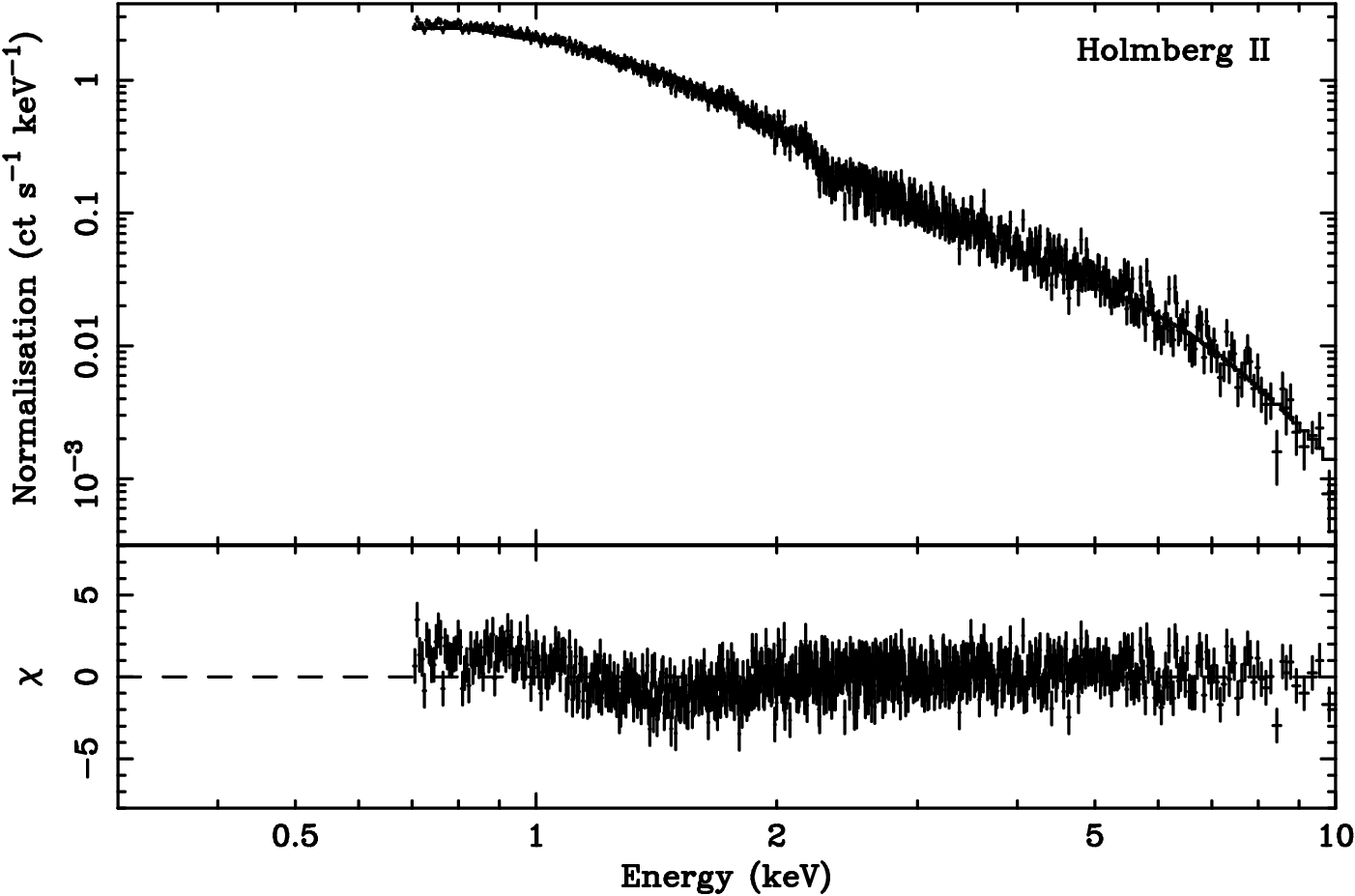}}}\vspace*{0.2cm}
\scalebox{0.35}{{\includegraphics[angle=270]{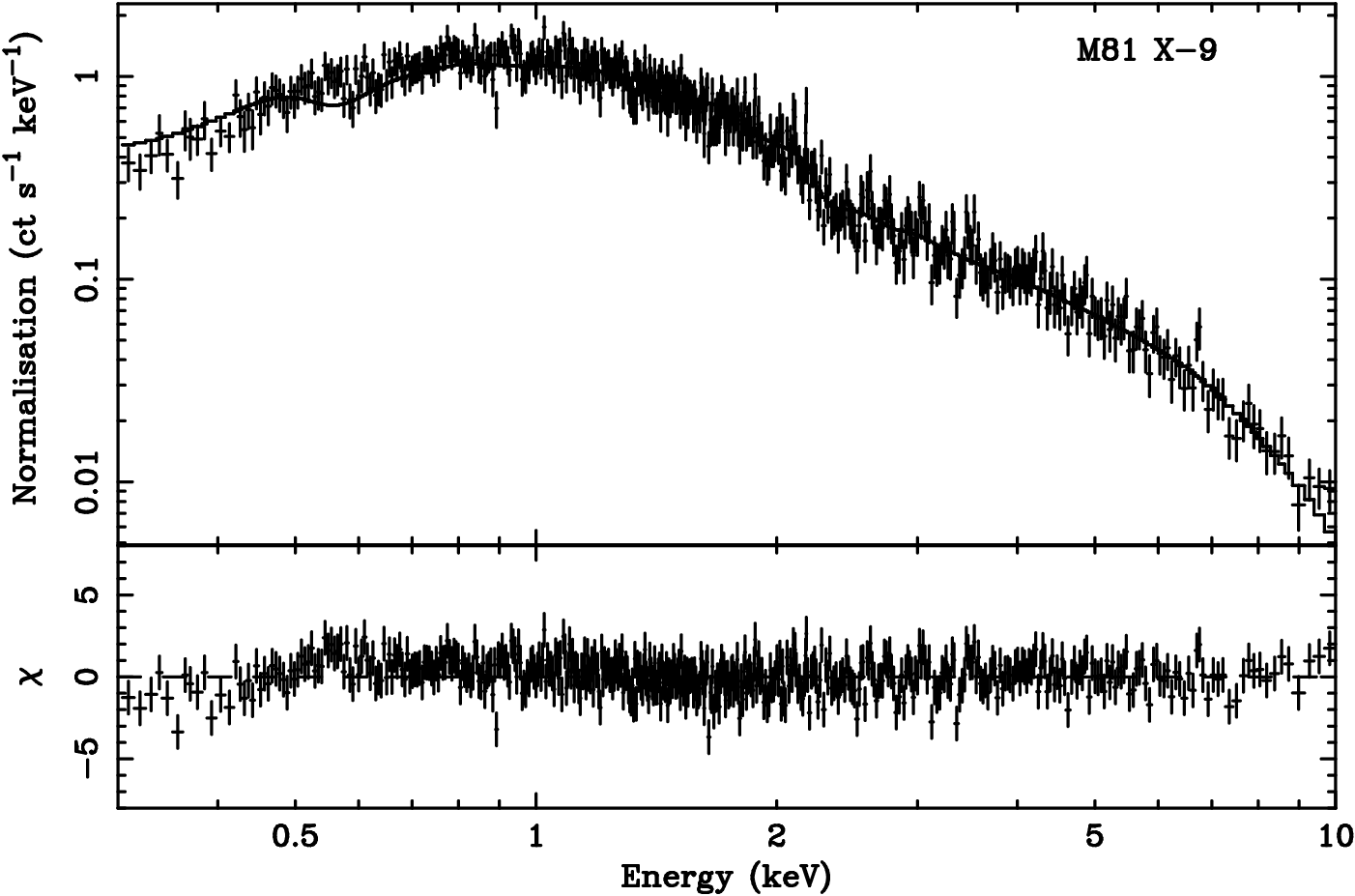}}}
\hspace*{0.5cm}
\scalebox{0.35}{{\includegraphics[angle=270]{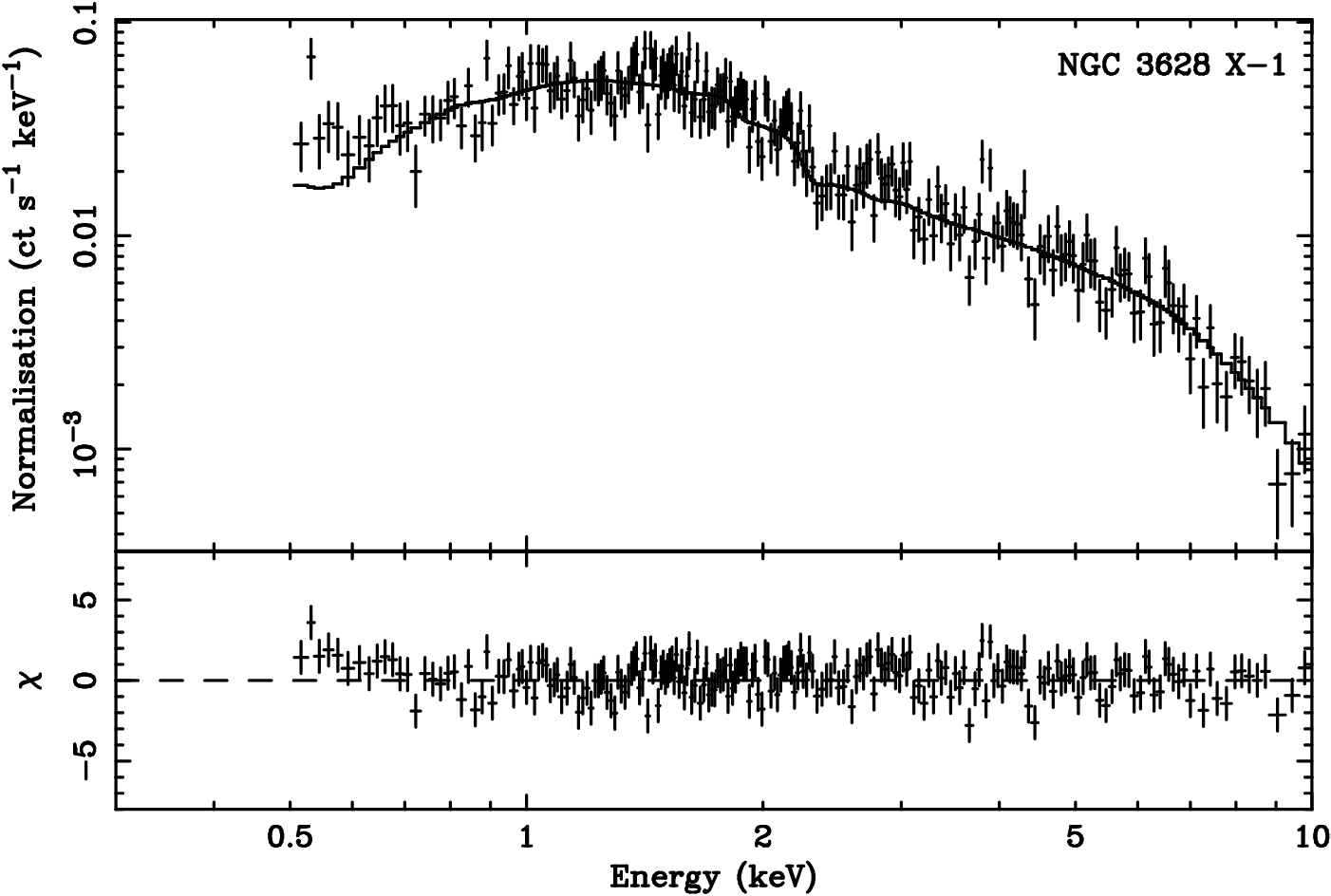}}}
\caption{\label{spec1}EPIC pn count rate spectra and \delchi
residuals for a simple power-law continuum model fit for each of the
ULXs in the sample.}
\end{center}
\end{figure*}
%=======================================================================
\begin{figure*}
\begin{center}
\scalebox{0.35}{{\includegraphics[angle=270]{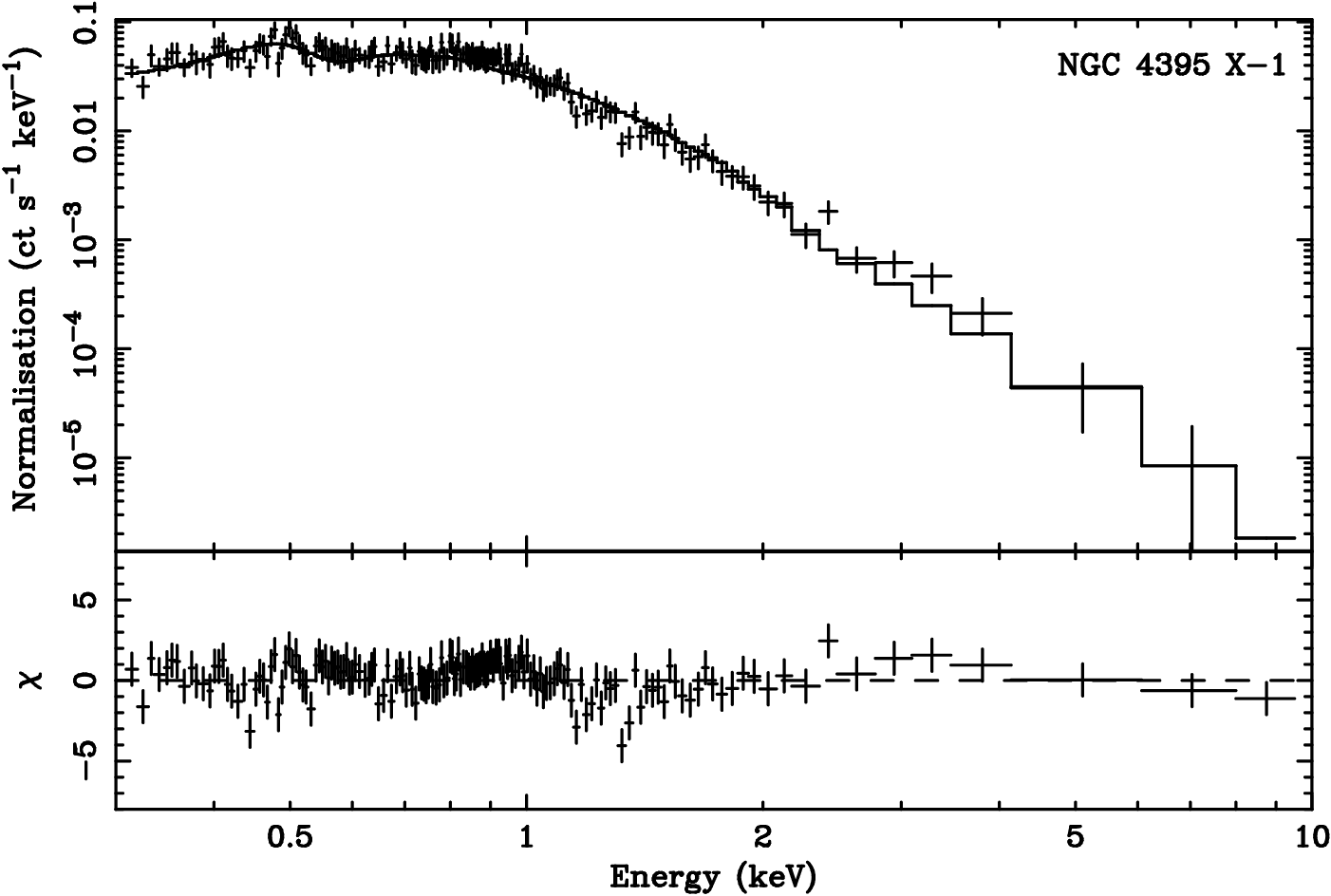}}}
\scalebox{0.35}{{\includegraphics[angle=270]{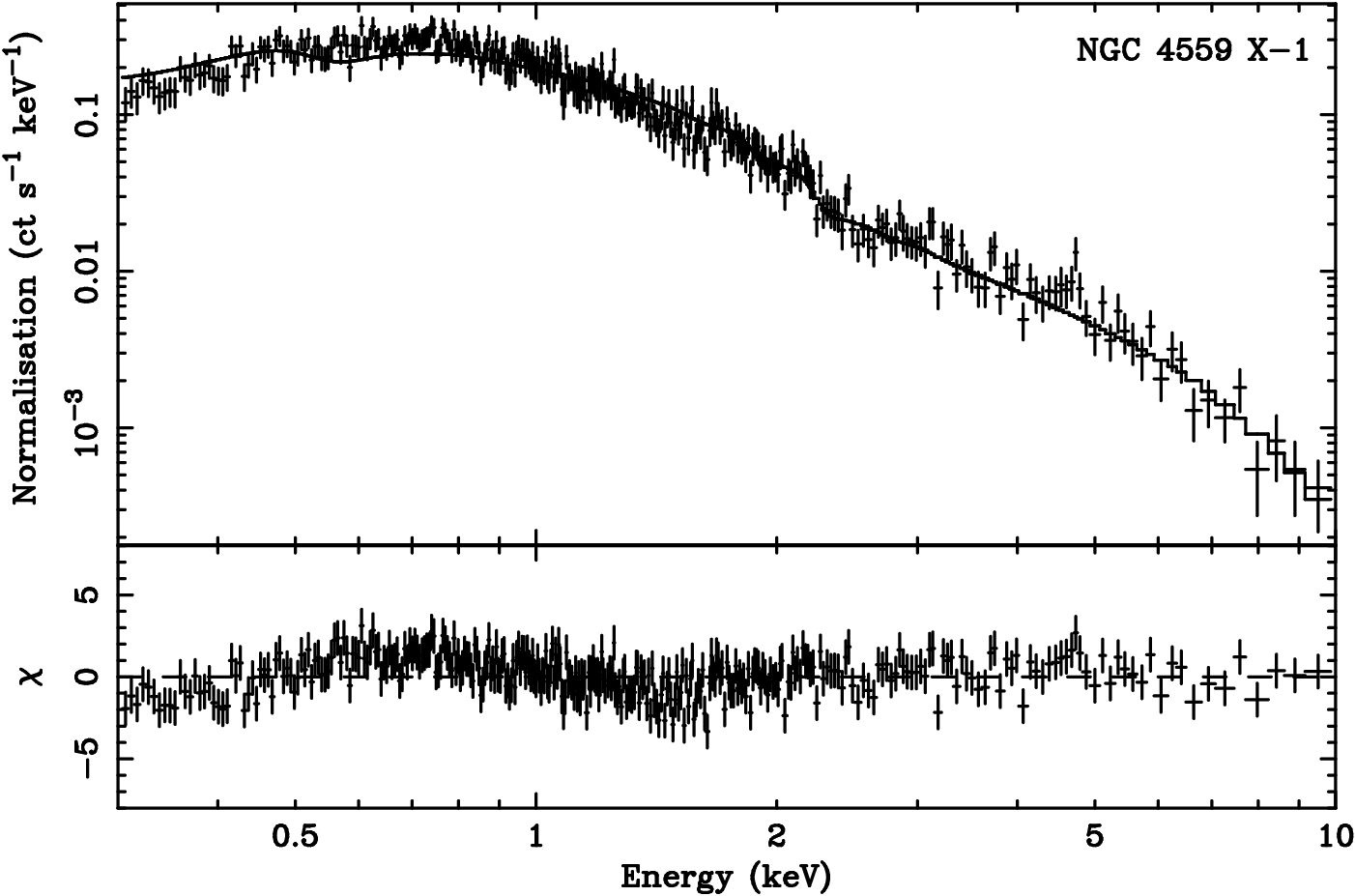}}}\vspace*{0.2cm}
\scalebox{0.35}{{\includegraphics[angle=270]{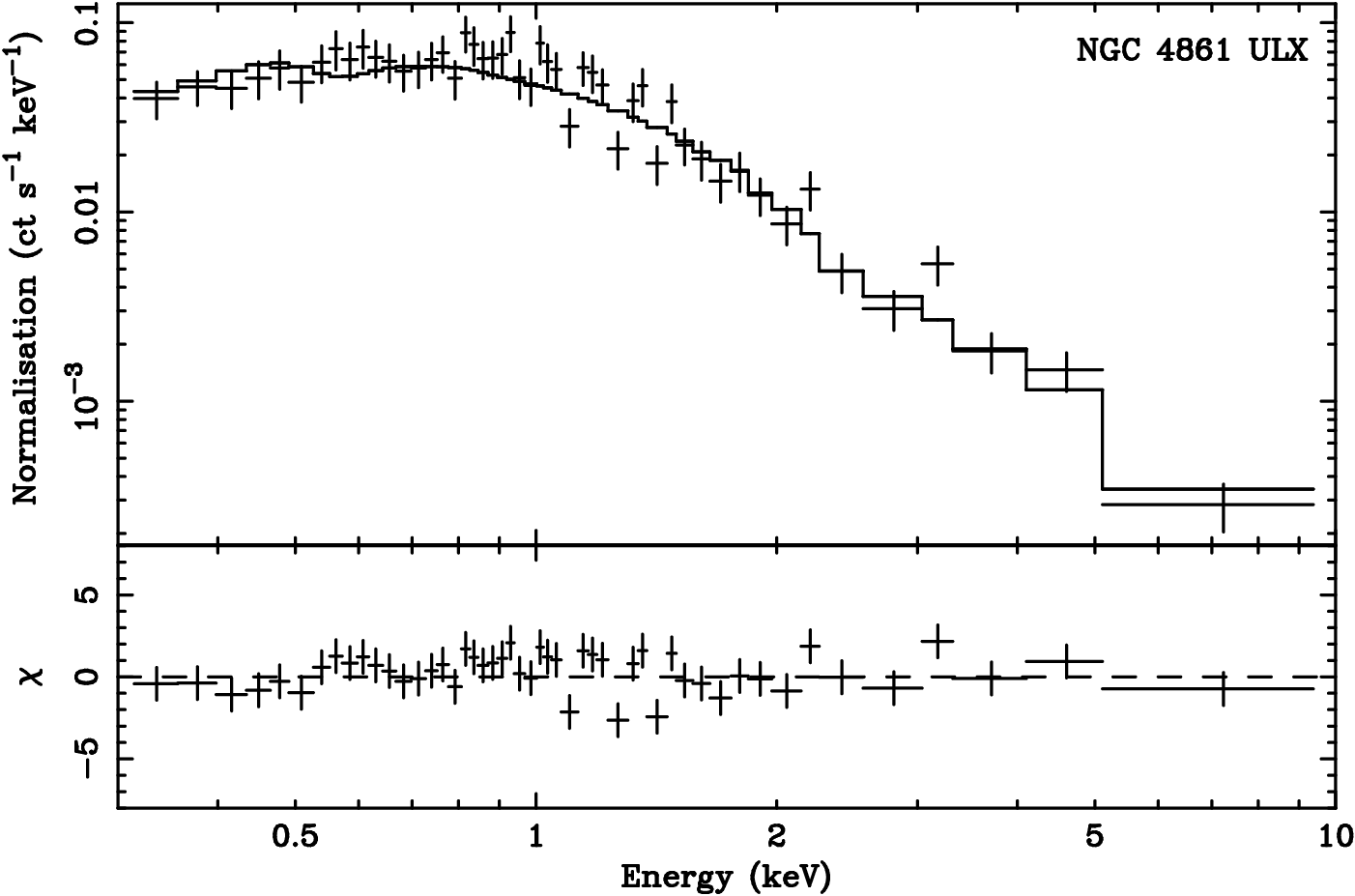}}}
\scalebox{0.35}{{\includegraphics[angle=270]{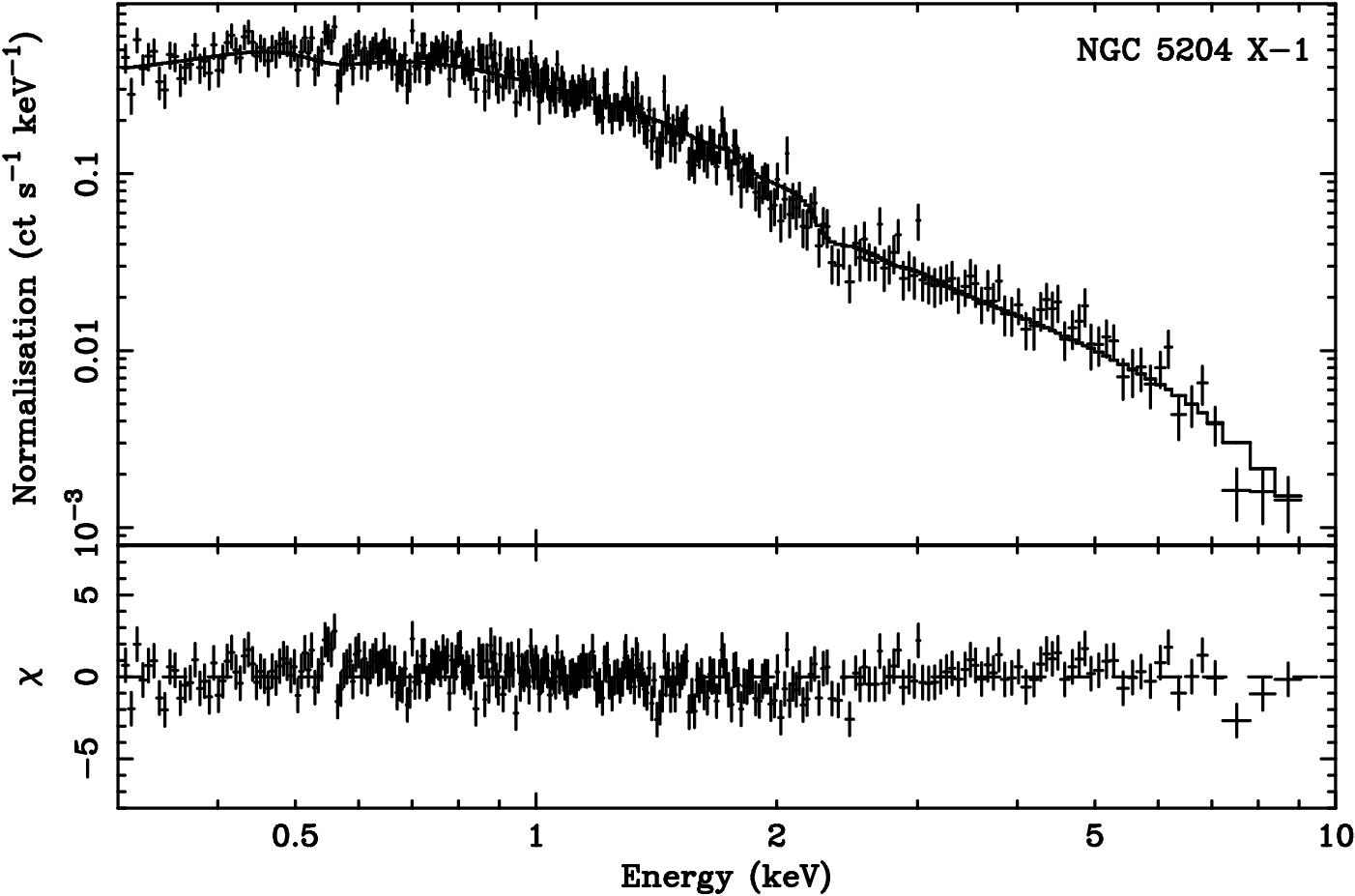}}}\vspace*{0.2cm}
\hspace*{4.5cm}
\scalebox{0.35}{{\includegraphics[angle=270]{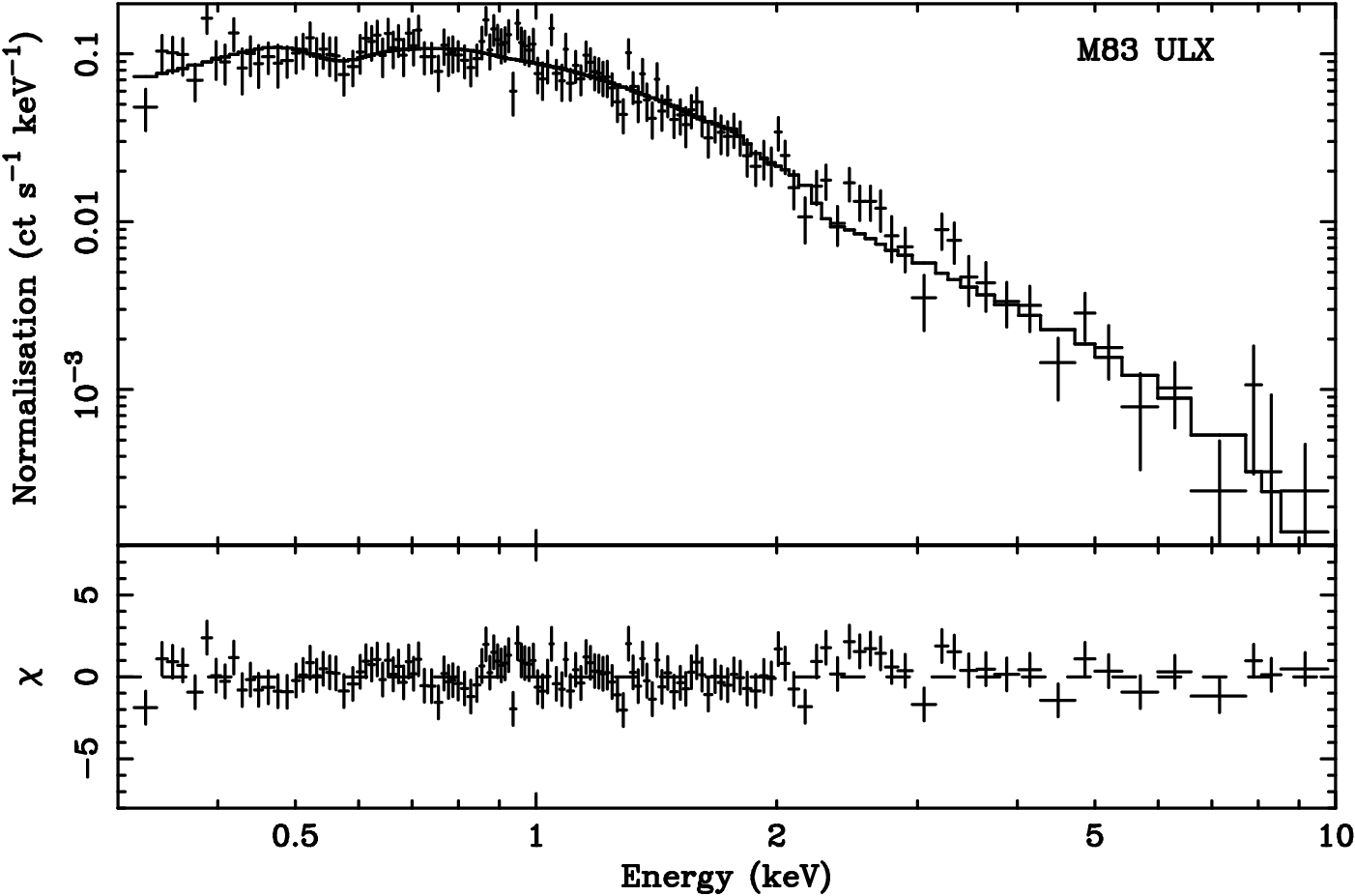}}}
\contcaption{}
\end{center}
\end{figure*}
%=======================================================================
\begin{table*}
\begin{center}
\small{ \caption{\label{single}Single component spectral fits}
\begin{tabular}{lcccc}
\hline
{\sc wa*wa*po}      &\nh$^a$        	&$\Gamma$$^b$       	&A$_{\rm P}$$^c$      	&\chisq/dof\\
\hline
NGC 55 ULX      &3.50$\pm0.08$      	&3.38$\pm0.03$      	&3.09$\pm0.08$      	&1128.9/830\\
M33 X-8         &1.94$\pm0.06$      	&2.28$\pm0.02$      	&6.02$\pm0.12$      	&1790.1/1115\\
NGC 1313 X-1    &1.05$\pm0.11$      	&1.84$^{+0.04}_{-0.03}$ &0.55$\pm0.02$      	&{\bf 723.0/675}\\
NGC 1313 X-2    &1.68$\pm0.02$      	&2.27$^{+0.08}_{-0.07}$ &0.31$\pm0.02$      	&{\bf 261.7/257}\\
NGC 2403 X-1    &3.78$\pm0.13$      	&2.40$\pm0.03$      	&0.65$\pm0.02$      	&1130.2/851\\
\hoii		&1.53$\pm0.04$ 		&2.76$\pm0.02$   	&3.23$\pm0.04$      	&1690.6/1314\\
M81 X-9         &1.57$\pm0.08$      	&1.89$\pm0.03$      	&1.92$^{+0.06}_{-0.05}$ &938.1/864\\
NGC 3628 X-1    &2.77$\pm0.03$      	&1.45$\pm0.05$      	&0.10$\pm0.01$      	&459.8/406\\
NGC 4395 X-1    &2.73$^{+0.17}_{-0.16}$ &4.37$^{+0.12}_{-0.11}$ &0.09$\pm0.01$      	&540.8/368\\
NGC 4559 X-1    &1.15$\pm0.08$      	&2.38$^{+0.05}_{-0.04}$ &0.27$\pm0.01$      	&752/599\\
NGC 4861 ULX    &1.24$^{+0.31}_{-0.30}$ &2.49$^{+0.17}_{-0.15}$ &0.09$\pm0.01$      	&{\bf 94.0/74}\\
NGC 5204 X-1    &0.50$\pm0.08$      	&2.10$^{+0.05}_{-0.04}$	&0.39$^{+0.02}_{-0.01}$ &{\bf 572.3/529}\\
M83 ULX         &1.22$^{+0.21}_{-0.20}$ &2.51$^{+0.11}_{-0.10}$ &0.14$\pm0.01$      	&{\bf 199.2/207}\\
\hline
{\sc wa*wa*diskbb}&\nh$^b$               &$kT_{\rm in}$$^d$      &A$_{\rm MCD}$$^e$      		&\chisq/dof\\
\hline
NGC 55 ULX      &0.37                    &0.60                   &1.62                   		&1796.3/830 \\
M33 X-8         &0			 &1.08$\pm0.01$          &0.62$^{+0.02}_{-0.01}$ 		&1538.7/1115 \\
NGC 1313 X-1    &0                       &1.32                   &0.04                   		&1409.6/675 \\
NGC 1313 X-2    &$<0.02$                 &0.93$\pm0.04$          &0.05$\pm0.01$          		&443.3/257 \\
NGC 2403 X-1    &1.20$^{+0.08}_{-0.07}$  &1.04$\pm0.02$          &0.06$\pm0.01$          		&923.7/851 \\
\hoii    	&0           		 &0.59                   &2.74                   		&8646.7/131\\
M81 X-9         &0			 &1.42                   &0.09                   		&1758.6/864\\
NGC 3628 X-1    &1.04$\pm0.02$           &2.27$^{+0.13}_{-0.12}$ &(1.38$^{+0.28}_{-0.24}) \times 10^{-3}$&466.6/406\\
NGC 4395 X-1    &0.13$\pm0.09$           &0.32$\pm0.01$          &0.81$^{+0.16}_{-0.13}$ 		&693.0/368\\
NGC 4559 X-1    &0                       &0.69                   &0.15                   		&1651.4/599\\
NGC 4861 ULX    &0                       &0.64                   &0.06                   		&166.9/74\\
NGC 5204 X-1    &0                       &0.66                   &0.33                   		&1650.1/529\\
M83 ULX         &$<0.01$                 &0.64$\pm0.03$          &0.11$\pm0.02$          		&310.7/207\\
\hline
\end{tabular}
\begin{minipage}[t]{4in}
{\sc Notes}: Models are abbreviated to {\sc xspec} syntax: {\sc
wa}--absorption components for the Galactic value and external
absorption; {\sc po}--power-law continuum; {\sc diskbb}--MCD.  $^a$
External absorption column ($10^{21} \atpcm$), $^b$ {\sc po} photon
index, $^c$ {\sc po} normalisation ($10^{-3}~\xspnorm$ at 1 keV ),
$^d$ Inner disc temperature (keV). $^e$ {\sc diskbb} normalisation
(((R$_{\rm in}$/km)/(D/10 kpc))$^2$ $\cos \Theta$; where R$_{\rm
in}$--inner disc radius, D--distance to source, $\Theta$--inclination
angle of the disc). Where the reduced $\chi^2$ value
(i.e. ${{\chi^2}\over{\rm dof}}$) exceeded 2, errors are not shown.
Statistically acceptable fits are highlighted in bold.
\end{minipage}
}
\end{center}
\end{table*}
%=======================================================================
%=======================================================================

\subsection{Power-law + MCD model}

\begin{table*}
\begin{center}
\small{ \caption{\label{twocomp}Two component spectral fits}
\begin{tabular}{lcccccccc}
\hline
{\sc wa*wa*(po+diskbb)}$^1$&\nh$^a$     &$\Gamma$$^b$       	&A$_{\rm P}$$^c$        &$kT_{\rm in}$$^d$        &A$_{\rm MCD}$$^e$($\times 10^{2}$)   &\chisq/dof        &f$_{\rm X_{\rm PO}}$$^{f}$ &f$_{\rm X_{\rm MCD}}$$^{g}$\\
\hline
NGC 55 ULX      &3.50$^{+0.42}_{-0.09}$ &3.38$\pm0.03$      	&3.09$^{+0.07}_{-0.08}$ &$< 0.16$                 &$< 9.71 \times 10^4$       		&1128.8/828        &99.9           &0.1\\
M33 X-8         &4.43$^{+0.25}_{-0.06}$ &2.52$^{+0.02}_{-0.01}$ &8.51$^{+0.20}_{-0.12}$	&0.09$\pm0.002$           &3312$^{+1694}_{-1055}$     		&1527.0/1113	   &92.7           &7.3\\
NGC 1313 X-1    &2.37$^{+0.52}_{-0.48}$ &1.77$\pm0.06$      	&0.51$\pm0.04$          &0.19$\pm0.02$            &0.81$^{+1.39}_{-0.55}$    		&{\bf 664.5/673}   &90.9           &9.1\\
NGC 1313 X-2    &1.80$^{+0.76}_{-0.42}$ &2.11$^{+0.15}_{-0.17}$ &0.25$^{+0.05}_{-0.06}$ &0.27$^{+0.10}_{-0.09}$   &0.03$^{+0.27}_{-0.02}$ 		&{\bf 255.0/255}   &91.7           &8.3\\
NGC 2403 X-1    &7.18$^{+0.04}_{-0.05}$ &2.68$^{+0.02}_{-0.05}$ &0.98$\pm0.03$          &0.09$^{+0.002}_{-0.004}$ &527$^{+644}_{-224}$    		&1028.5/849        &93.6           &6.4\\
\hoii		&1.74$^{+0.10}_{-0.09}$ &2.63$\pm0.03$		&2.81$\pm0.09$          &0.19$\pm0.01$            &211$^{+1.23}_{-0.78}$  		&1453.1/1312       &89.7           &10.3\\
M81 X-9     	&2.17$^{+0.43}_{-0.23}$ &1.81$^{+0.05}_{-0.04}$ &1.74$^{+0.14}_{-0.10}$ &0.20$\pm0.03$            &1.05$^{+2.78}_{-0.61}$ 		&{\bf 877.2/862}   &94.0           &6.0\\
NGC 3628 X-1    &4.26$^{+0.77}_{-0.70}$ &1.58$^{+0.07}_{-0.04}$ &0.12$\pm0.01$          &0.08$\pm0.01$            &127$^{+402}_{-96}$         		&{\bf 424.6/404}   &98.1           &1.9\\
NGC 4395 X-1    &1.96$^{+0.26}_{-0.40}$ &3.72$^{+0.18}_{-0.27}$ &0.05$\pm0.01$          &0.18$\pm0.02$            &0.11$^{+0.09}_{-0.05}$ 		&492.2/366         &71.6           &28.4\\
NGC 4559 X-1    &2.33$^{+0.70}_{-0.20}$ &2.23$^{+0.07}_{-0.05}$ &0.24$^{+0.03}_{-0.01}$ &0.14$\pm0.01$            &2.74$^{+5.64}_{-0.96}$ 		&{\bf 597.2/597}   &81.5           &18.5\\
NGC 4861 ULX    &1.75$^{+1.13}_{-0.69}$ &2.24$^{+0.23}_{-0.24}$ &0.07$\pm0.02$          &0.18$^{+0.08}_{-0.04}$   &0.13$^{+0.03}_{-0.12}$ 		&{\bf 84.4/72}     &82.4           &17.6\\
NGC 5204 X-1    &0.68$^{+0.22}_{-0.18}$ &1.91$^{+0.07}_{-0.08}$ &0.31$\pm0.03$          &0.22$^{+0.04}_{-0.03}$   &0.12$^{+0.21}_{-0.07}$ 		&{\bf 529.9/527}   &88.2           &11.8\\
M83 ULX     	&1.28$^{+0.54}_{-0.37}$ &2.47$^{+0.14}_{-0.17}$ &0.14$^{+0.02}_{-0.03}$ &$[\sim 0.2]~^k$          &$<0.98$                    		&{\bf 198.6/205}   &96.6           &3.4\\
\hline
{\sc wa*wa*(po+diskbb)}$^2$&\nh$^a$     &$\Gamma$$^b$           &A$_P$$^c$          	&$kT_{\rm in}$$^d$        &A$_{\rm MCD}$$^e$($\times 10^{-3}$)  &\chisq/dof        &f$_{\rm X_{\rm PO}}$$^{f}$ &f$_{\rm X_{\rm MCD}}$$^{g}$\\
\hline
NGC 55 ULX      &4.32$^{+0.24}_{-0.34}$ &4.31$^{+0.19}_{-0.27}$ &3.29$^{+0.21}_{-0.27}$ &0.86$\pm0.03$            &170.47$^{+35.51}_{-40.63}$     	&957.6/828         &67.6           &32.4\\
M33 X-8         &1.42$^{+0.21}_{-0.18}$ &2.49$^{+0.18}_{-0.14}$ &3.15$^{+0.31}_{-0.32}$ &1.18$\pm0.06$            &264.55$^{+53.06}_{-43.13}$     	&{\bf 1187.0/1113} &48.8           &51.2\\
NGC 1313 X-1    &--                 	&--                 	&--                 	&--                       &--                 			&--                &--             &--\\
NGC 1313 X-2    &2.52$^{+0.97}_{-0.62}$ &2.93$^{+0.75}_{-0.47}$ &0.36$^{+0.09}_{-0.05}$ &2.65$^{+1.06}_{-0.78}$   &$<$2.19                    		&{\bf 255.9/255}   &64.4           &35.6\\
NGC 2403 X-1    &4.77$^{+1.01}_{-0.99}$ &4.05$^{+0.70}_{-0.75}$ &0.55$^{+0.20}_{-0.15}$ &1.16$^{+0.04}_{-0.05}$   &36.93$^{+10.14}_{-7.95}$       	&{\bf 847.7/849}   &27.3           &72.7\\
\hoii		&2.34$^{+0.13}_{-0.12}$ &3.45$^{+0.11}_{-0.10}$ &3.78$^{+0.11}_{-0.10}$&1.79$^{+0.10}_{-0.09}$	  &10.33$^{+3.33}_{-2.70}$  		&1508.8/1312 	   &74.5 	   &25.5\\
M81 X-9         &--                 	&--                 	&--                 	&--                       &--                 			&--                &--             &--\\
NGC 3628 X-1    &1.43$^{+0.47}_{-0.26}$ &$<1.18$        	&$<0.04$            	&1.40$\pm0.21$            &4.68$^{+1.65}_{-2.31}$         	&{\bf 439.5/404}   &53.4           &46.6\\
NGC 4395 X-1    &--                 	&--                 	&--                 	&--                       &--                 			&--                &--             &--\\
NGC 4559 X-1    &3.35$^{+0.46}_{-0.41}$ &4.47$^{+0.37}_{-0.34}$ &0.41$^{+0.05}_{-0.04}$ &1.61$^{+0.13}_{-0.12}$   &4.04$^{+1.64}_{-1.59}$             	&{\bf 637.0/597}   &47.3           &52.7\\
NGC 4861 ULX    &3.33$^{+1.75}_{-1.30}$ &4.36$^{+1.34}_{-1.08}$ &0.13$^{+0.08}_{-0.04}$ &1.60$^{+0.60}_{-0.32}$   &1.09$^{+1.34}_{-0.87}$             	&{\bf 85.6/72}     &51.9           &48.1\\
NGC 5204 X-1    &1.64$^{+0.37}_{-0.33}$ &3.33$^{+0.34}_{-0.32}$ &0.44$^{+0.05}_{-0.03}$ &2.23$^{+0.32}_{-0.26}$   &1.96$^{+1.17}_{-0.87}$             	&{\bf 525.5/527}   &50.3           &49.7\\
M83 ULX         &1.37$^{+0.92}_{-0.45}$ &2.75$^{+0.94}_{-0.45}$ &0.14$\pm0.02$          &$<4.03$                                    &$<$4.53            &{\bf 198.1/205}   &82.9           &17.1\\
\hline
{\sc wa*wa*(bb+diskbb)}&\nh$^a$         &$kT$$^h$               &A$_{\rm B}$$^i$        &$kT_{\rm in}$$^d$        &A$_{\rm MCD}$$^e$($\times 10^{-3}$)  &\chisq/dof        &f$_{\rm X_{\rm BB}}$$^{j}$ &f$_{\rm X_{\rm MCD}}$$^{g}$\\
\hline
NGC 55 ULX      &1.43$^{+0.16}_{-0.17}$ &0.20$\pm0.01$          &3.35$^{+0.33}_{-0.31}$ &0.81$\pm0.01$            &343$^{+53}_{-47}$              	&{\bf 884.1/828}   &36.4           &63.6\\
M33 X-8         &0.16$^{+0.10}_{-0.09}$ &0.27$\pm0.02$          &3.09$^{+0.31}_{-0.30}$ &1.26$^{+0.02}_{-0.03}$   &303$^{+37}_{-16}$              	&1216.7/1113       &11.7           &88.3\\
NGC 1313 X-1    &0.60$^{+0.14}_{-0.25}$ &0.25$^{+0.02}_{-0.01}$ &0.86$^{+0.11}_{-0.10}$ &2.20$\pm0.10$            &5.07$^{+0.31}_{-0.16}$         	&{\bf 666.5/673}   &17.7           &82.3\\
NGC 1313 X-2    &$<0.54$                &0.27$^{+0.04}_{-0.02}$ &0.41$^{+0.07}_{-0.06}$ &1.71$^{+0.15}_{-0.16}$   &4.02$^{+0.49}_{-1.36}$         	&{\bf 257.1/255}   &27.7           &72.3\\
NGC 2403 X-1    &2.02$\pm0.30$          &0.21$^{+0.03}_{-0.02}$ &0.38$^{+0.14}_{-0.09}$ &1.12$^{+0.04}_{-0.03}$   &46.63$\pm3.42$                 	&{\bf 830.0/849}   &8.7            &91.3\\
\hoii		&$<0.01$        	&0.23$\pm0.002$         &4.37$^{+0.06}_{-0.07}$ &1.28$\pm0.02$            &77.69$^{+6.44}_{-6.05}$        	&1677.9/1312       &41.4           &58.6\\
M81 X-9         &0.64$^{+0.14}_{-0.10}$ &0.27$\pm0.01$          &2.56$\pm0.16$          &2.21$^{+0.12}_{-0.11}$   &15.93$^{+3.34}_{-2.83}$        	&{\bf 922.3/862}   &17.5           &82.5\\
NGC 3628 X-1    &0.71$^{+0.28}_{-0.24}$ &0.60$^{+0.05}_{-0.06}$ &0.16$^{+0.04}_{-0.06}$ &4.32$^{+0.74}_{-0.87}$   &0.13$^{+0.17}_{-0.07}$         	&{\bf 429.4/404}   &17.7           &82.3\\
NGC 4395 X-1    &0.34$^{+0.19}_{-0.17}$ &0.17$\pm0.01$          &0.15$\pm0.02$          &0.61$\pm0.05$            &17.66$^{+9.56}_{-6.26}$        	&464.5/366         &65.5           &34.5\\
NGC 4559 X-1    &0.80$^{+0.17}_{-0.11}$ &0.17$\pm0.01$          &0.61$^{+0.11}_{-0.08}$ &1.33$^{+0.07}_{-0.06}$   &9.94$^{+2.29}_{-1.82}$         	&{\bf 604.9/597}   &30.6           &69.4\\
NGC 4861 ULX    &0.46$^{+0.05}_{-0.03}$ &0.20$\pm0.03$          &0.15$^{+0.09}_{-0.04}$ &1.37$^{+0.20}_{-0.08}$   &2.38$^{+1.89}_{-1.13}$         	&{\bf 77.7/72}     &34.1           &65.9\\
NGC 5204 X-1    &$<0.03$                &0.20$\pm0.01$          &0.67$\pm0.03$          &1.69$^{+0.10}_{-0.09}$   &7.37$^{+1.69}_{-1.40}$         	&{\bf 527.8/527}   &28.0           &72.0\\
M83 ULX         &$<0.37$            	&0.21$^{+0.02}_{-0.03}$ &0.16$^{+0.05}_{-0.02}$ &1.10$^{+0.13}_{-0.05}$   &9.28$^{+2.76}_{-3.55}$             	&{\bf 193.2/205}   &27.8           &72.2\\
\hline
\end{tabular}
\begin{minipage}[t]{6.6in}
{\sc Notes}: Models are abbreviated to {\sc xspec} syntax: {\sc wa},
{\sc po} and {\sc diskbb} as before, {\sc bb}--blackbody
continuum. Model fitted with a cool$^1$ or hot$^2$ {\sc diskbb}
component. $^a$ External absorption column ($10^{21} \atpcm$), $^b$
{\sc po} photon index, $^c$ {\sc po} normalisation ($10^{-3}~\xspnorm$
at 1 keV), $^d$ inner disc temperature (keV).  $^e$ {\sc diskbb}
normalisation as before (((R$_{\rm in}$/km)/(D/10 kpc))$^2$ $\cos
\Theta$), $^f$ fraction of the total flux (0.3--10 keV) in the {\sc
po} component, $^g$ fraction of the total flux (0.3--10 keV) in the
{\sc diskbb} component, $^h$ blackbody temperature (keV), $^i$
blackbody normalisation, ($10^{-5}~$L$_{39}$/D$^{2}_{10}$; where
L$_{39}$--source luminosity in $10^{39} \ergsec$, D$_{10}$--source
distance in 10 kpc). $^j$ Fraction of the total flux (0.3--10 keV) in
the {\sc bb} component.  $^k$ Unconstrained at the 90 per cent
confidence level.  In the three cases in the central portion of the
table where no fit is shown, the $\chi^2$-minimisation always found
the minimum describing the IMBH model.  Statistically acceptable fits
are highlighted in bold. (NB. In cases where the spectral fitting was
not performed over the 0.3--10 keV energy range, we created a dummy
response matrix in {\sc xspec} (see Sec. 4.5), to determine the
fraction of the total flux in each spectral component over 0.3--10
keV, consistent with the other datasets).
\end{minipage}
}
\end{center}
\end{table*}
%=======================================================================

As simple spectral models were inadequate, we proceeded to fit the
data with the two component model generally employed for BHB systems,
namely the combination of a power-law plus a MCD component.  More
specifically, we began by testing the standard IMBH model, \ie a cool
accretion disc plus power-law continuum model. This spectral
description improved on the simple models by providing acceptable fits
to 8 sources (NGC 1313 X-1 \& X-2, M81 X-9, NGC 3628 X-1, NGC 4559
X-1, NGC 4861 ULX, NGC 5204 X-1, M83 ULX), with 0.1 keV $< kT_{\rm in}
< 0.3$ keV and $1.6 < \Gamma < 2.5$ (Table \ref{twocomp}).  As with
previously published work, these disc temperatures are broadly
consistent with IMBHs of around $\sim$1000-\Msun in size, though the
power-law slopes are puzzlingly shallow for what are supposed HS (or
VH state) sources (see \citealt{roberts05} for further discussion). In
all cases the 0.3--10 keV flux is dominated by the power-law
component.

We then attempted to fit the X-ray spectra with the non-standard model
\ie with the power-law component dominant at soft energies. This also
provided statistically-acceptable fits to 8 ULXs (M33 X-8, NGC 1313
X-2, NGC 2403 X-1, NGC 3628 X-1, NGC 4559 X-1, NGC 4861 ULX, NGC 5204
X-1, M83 ULX), with disc temperatures of 1.2 keV $< kT_{\rm in} < 2.2$
keV (excluding M 83 for which the temperature is unconstrained below
$\sim$4 keV) and $2.5 < \Gamma < 4.5$ (except NGC 3628 X-1, for which
$\Gamma$ is unconstrained below $\sim$1.2) (Table \ref{twocomp}).  The
balance in 0.3--10 keV flux between the two components was much more
varied for this model, though most sources showed a relatively
equitable balance (ratios of no more than 3:1 in either direction, and
$\sim$1:1 in five cases).

In total, the power-law + MCD combination could not be rejected at $>$
95 per cent confidence for 10/13 sources in the sample. Two of these
sources were unambiguously best fitted with the IMBH model (NGC 1313
X-1, M81 X-9) while two more were best fitted by the non-standard
model (M33 X-8, NGC 2403 X-1). Statistically acceptable fits existed
{\it for both models\/} for the remaining six sources.  Three sources
in this sample (NGC 55 ULX, \hoii and NGC 4395) rejected both models
at $>95$ percent confidence, although a comparison of the fits show
that the spectrum of the NGC 55 ULX appeared to fit much better to the
non-standard model, whilst Ho II X-1 and NGC 4395 X-1 much preferred
the IMBH model.

This sample of high quality ULX datasets demonstrates spectral
ambiguity for six sources (NGC 1313 X-2, NGC 3628 X-1, NGC 4559 X-1,
NGC 4861 ULX, NGC 5204 X-1, M83 ULX). As \citet{roberts05} suggest, a
key discriminator between the models could be a test for spectral
curvature at high energies ($> 2$ keV).  Curvature is not expected in
the IMBH model where the power-law component is dominant above 2 keV,
whereas sources best-fitted by the non-standard model have the curved
MCD component dominant at these hard X-ray energies.  We therefore
attempted to fit the high energy (2--10 keV) ULX spectra with a broken
power-law model.  For comparison we also fitted the 2--10 keV ULX
spectra with a single power-law component, as shown in Table
\ref{power}.  We did not include absorption in the fit in either case,
as single- and two-component fits generally limit absorption to $\la 4
\times 10^{21} \atpcm$, which should not strongly affect the data
above 2 keV.  When we do include absorption in these fits we actually
measure far higher columns in several cases, which is physically
unrealistic (and may be a result of the absorption compensating for
intrinsic curvature).  The results of the broken power-law spectral
fits, together with the statistical probability of the fit improvement
over the single power-law fits (using the F-test) are shown in
Table~\ref{broken}.

To demonstrate the validity of this method, the two sources with
unambiguous non-standard fits (M33 X-8 and NGC 2403 X-1), plus NGC 55
ULX which clearly prefers this model, show unacceptable power-law fits
that are made acceptable by the inclusion of a break.  In these cases
the improvement is highly statistically significant ($> 9 \sigma$
improvement over a simple power-law fit according to the F-test).
However, these were the clearest-cut cases.  Of the ambiguous spectra,
three out of six showed evidence for curvature.  This was marginal in
the case of M83 ULX ($> 2 \sigma$ level), but more significant for NGC
4559 X-1 and NGC 5204 X-1 ($> 3 \sigma$), with the latter case having
an unacceptable power-law fit.  Most interestingly, though, one of the
IMBH-fit sources (NGC 1313 X-1) shows evidence for curvature ($> 3
\sigma$ improvement), whilst Ho II X-1 also shows a significant break
($>4 \sigma$ improvement).  In total 8/13 sources show at least
marginal evidence for curvature at the high energy end of their \xmmn
spectrum. This is strongly suggestive that the majority of sources are
{\it not\/} dominated by a power-law continuum at these energies, such
as one might expect to see if the X-ray emission arises from the
optically thin, hot corona assumed in the IMBH model.

%=======================================================================
\begin{table*}
\begin{center}
\small{ \caption{\label{power}Power-law spectral fits (2--10 keV)}
\begin{tabular}{lccc}
\hline
{\sc wa*wa*po}  &$\Gamma$$^a$       		&A$_{\rm P}$$^b$         	&\chisq/dof\\
\hline			
NGC 55 ULX 	&3.58$^{+0.06}_{-0.07}$ 	&3.57$\pm0.25$		        &459.3/326\\
M33 X-8    	&2.59$\pm0.04$      		&8.90$^{+0.41}_{-0.39}$         &663.1/548\\
NGC 1313 X-1    &1.70$^{+0.05}_{-0.07}$ 	&0.44$^{+0.03}_{-0.04}$     	&{\bf 263.7/259}\\
NGC 1313 X-2    &2.19$^{+0.12}_{-0.16}$ 	&0.27$^{+0.04}_{-0.05}$         &{\bf 60.2/70}\\
NGC 2403 X-1	&2.63$^{+4.95}_{-0.07}$ 	&0.82$^{+0.05}_{-0.06}$     	&467.3/340\\
\hoii  		&2.61$\pm0.03$      		&2.65$\pm0.10$		    	&{\bf 805.7/824}\\
M81 X-9    	&1.78$\pm0.05$ 			&1.59$\pm0.10$         		&{\bf 360.4/355}\\
NGC 3628 X-1    &1.50$^{+0.06}_{-0.08}$ 	&0.10$\pm0.01$		        &{\bf 219.5/211}\\
NGC 4395 X-1   	&4.48$^{+0.31}_{-0.45}$ 	&0.12$\pm0.04$                  &{\bf 33.1/28}\\
NGC 4559 X-1  	&2.28$^{+0.08}_{-0.10}$ 	&0.24$^{+0.02}_{-0.03}$     	&{\bf 171.3/155}\\
NGC 4861 ULX    &2.50$\pm0.38$ 			&0.09$^{+0.06}_{-0.04}$         &{\bf 12.3/8}\\
NGC 5204 X-1    &1.89$\pm0.09$ 			&0.29$^{+0.04}_{-0.03}$     	&160.8/133\\
M83 ULX    	&2.67$^{+0.21}_{-0.28}$ 	&0.17$\pm0.05$	 	        &{\bf 45.5/39}\\
\hline
\end{tabular}

\begin{minipage}[t]{2.7in}
{\sc Notes}: Models are abbreviated to {\sc xspec} syntax: {\sc po} as
before.  $^a$ {\sc po} photon index, $^b$ {\sc po} normalisation
($10^{-3}~\xspnorm$ at 1 keV). Statistically acceptable fits are
highlighted in bold.
\end{minipage}
}
\end{center}
\end{table*}

%=======================================================================

\begin{table*}
\begin{center}
\small{ \caption{\label{broken}Broken power-law spectral fits (2--10 keV)}
\begin{tabular}{lccccccc}
\hline
{\sc wa*wa*bknpo}   	  &$\Gamma_1$$^a$         &$E_{\rm break}$$^b$        	&$\Gamma_2$$^c$         	&A$_{\rm BP}$$^d$        &\chisq/dof        &$\Delta$\chisq$^{e}$   	&1-P(F-test)$^{f}$\\
\hline			
NGC 55 ULX          	  &3.08$\pm0.11$ 	  &3.90$^{+0.20}_{-0.18}$	&5.34$^{+0.53}_{-0.41}$ 	&2.32$^{+0.26}_{-0.24}$  &{\bf 341.0/324}   &118.3       		&$>$99.9 \\
M33 X-8             	  &2.17$\pm0.09$  	  &3.94$^{+0.26}_{-0.21}$	&3.32$^{+0.19}_{-0.15}$ 	&6.07$^{+0.56}_{-0.52}$  &{\bf 516.1/546}   &147.0       		&$>$99.9 \\
NGC 1313 X-1       	  &1.55$^{+0.10}_{-0.09}$ &4.93$^{+1.37}_{-0.53}$ 	&2.16$^{+0.66}_{-0.25}$ 	&0.37$^{+0.05}_{-0.04}$  &{\bf 250.9/257}   &12.9        		&99.8 \\
NGC 1313 X-2        	  &-- 			  &-- 				&-- 				&-- 			 &-- 		    &--				&--\\
NGC 2403 X-1        	  &2.07$^{+0.10}_{-0.11}$ &4.00$^{+0.17}_{-0.15}$ 	&4.05$^{+0.33}_{-0.28}$ 	&0.49$\pm0.05$  	 &{\bf 325.4/338}   &141.9      		&$>$99.9 \\
\hoii  			  &2.55$\pm0.04$	  &5.31$^{+0.43}_{-0.56}$ 	&3.08$^{+0.25}_{-0.22}$ 	&2.50$^{+0.11}_{-0.12}$  &{\bf 786.1/822}   &19.6     		 	&$>$99.9  \\
M81 X-9             	  &-- 			  &-- 				&-- 				&-- 			 &--  		    &--				&--\\
NGC 3628 X-1        	  &-- 			  &-- 				&-- 				&-- 			 &--  		    &--				&--\\
NGC 4395 X-1        	  &2.77$^{+1.31}_{-2.58}$ &2.45$^{+0.93}_{-0.23}$	&4.92$^{+0.77}_{-0.60}$ 	&$<$0.09		 &{\bf 30.0/26}     &3.1			&72.1\\
NGC 4559 X-1        	  &2.07$^{+0.12}_{-0.14}$ &4.77$^{+0.65}_{-0.41}$ 	&3.14$^{+0.62}_{-0.43}$ 	&0.20$\pm0.03$           &{\bf 157.7/153}   &13.6       		&99.8 \\
NGC 4861 ULX        	  &--  &--		&--		&--  &--      &--	&-- \\
NGC 5204 X-1        	  &1.65$^{+0.10}_{-0.15}$ &4.92$^{+0.46}_{-0.41}$ 	&2.99$^{+0.65}_{-0.45}$ 	&0.23$^{+0.03}_{-0.04}$  &{\bf 138.3/131}   &22.5      		  	&$>$99.9 \\
M83 ULX             	  &$<$1.79 		  &2.41$^{+0.31}_{-0.15}$ 	&3.11$^{+0.46}_{-0.40}$ 	&$<0.08$                 &{\bf 37.5/37}     &8.0      			&97.2 \\
\hline
\end{tabular}

\begin{minipage}[t]{5.2in}
{\sc Notes}: Models are abbreviated to {\sc xspec} syntax:{\sc
bknpo}--broken power-law model. $^a$ Photon index below the break
energy, $^b$ break point for the energy (keV), $^c$ photon index above
the break energy, $^d$ {\sc bknpo} normalisation ($10^{-3}~\xspnorm$
at 1 keV), $^e$ \chisq improvement over a single {\sc po} fit, for two
extra degrees of freedom, $^f$ statistical probability (per cent) of
the fit improvement over a single {\sc po} fit. We only show the
results where we could constrain fits that showed the requisite
behaviour, i.e. $\Gamma_2 > \Gamma_1$ and $E_{\rm break}$ constrained
in the 2--10 keV range.. Statistically acceptable fits are highlighted
in bold.
\end{minipage}
}
\end{center}
\end{table*}

%=======================================================================

\subsection{Dual thermal models}

In the above analysis we determine that there is (at least marginal)
evidence of curvature above 2 keV in the majority of our sources,
arguing that a power-law continuum is not an adequate description of
this emission, while the hotter accretion disc present in the
alternate model may be more appropriate.  The situation below 2 keV is
less clear cut.  Though the alternate model uses a power-law continuum
to describe the soft component, this is modified to appear curved by
absorption.  As this power-law emission is difficult to understand
physically we have therefore explored a new variation of the alternate
model in which the soft component is also intrinsically curved (\ie it
does not rely solely on absorption to produce the observed curvature).
In this model we fitted a cool blackbody (BB) continuum to the soft
component, and a MCD to the hard component.  The physical motivation
for the soft component comes from the suggestion that optically-thick
outflowing winds from BHs accreting at or above the Eddington limit
could explain ultrasoft components in ULXs \citep{king03}.  Such winds
would appear as BB continua with temperatures of $\sim$0.1--0.3 keV.
One might then observe emission from both the accretion disc and the
wind given a favourable accretion disc geometry and viewing angle
(A. King, priv. comm.)\footnote{\citet{miller04a} argue that cool,
optically-thick outflows cannot explain the soft excess in ULXs as it
would be impossible to form powerful enough shocks to produce a
luminous non-thermal power-law continuum component in such systems.
If the hard component is not a power-law, but instead originates
directly from the accretion disc, then the requirement for shocks to
be present is removed and their argument is circumvented.}.

The results of applying this new model to the data are shown in Table
\ref{twocomp}.  It proved to be the most successful empirical
description of the data, providing statistically acceptable fits to 10
out of the 13 sources, with 0.15 keV $< kT < 0.3$ keV and 0.8 keV $<
kT_{\rm in} < 2.2$ keV (excluding NGC 3628 X-1 for which the
temperatures are much higher than shown in the other sources, \ie
$kT$$\sim$0.6 keV and $kT_{\rm in}$$\sim$4.3 keV).  This model
provided the only acceptable empirical fit to the ULX in NGC 55, and
all six ambiguous sources are adequately described by this model.
Curiously, this is also true for both IMBH sources (though it does
provide a notably worse fit than the IMBH model for M 81 X-9), and for
one of the two non-standard model sources (NGC 2403 X-1).  However,
this model could not provide statistically acceptable fits to the
other non-standard source (M 33 X-8) or to the final two ULXs (Ho II
X-1 \& NGC 4395 X-1, which also do not fit well with the power-law +
MCD combination).

%=======================================================================

\subsection{Why do \hoii and NGC 4395 X-1 not conform?}

%=======================================================================
\begin{table*}
\begin{center}
\footnotesize{\caption{\label{mekalpo}{\sc mekal} + power-law spectral fits}
\begin{tabular}{lcccccccc}
\hline
{\sc wa*wa*(mekal+po)}    &\nh$^a$          	&$kT$$^b$       &A$_{\rm M}$$^c$          	&$\Gamma$$^d$           &A$_{\rm P}$$^e$      	&\chisq/dof		&f$_{\rm X_{\rm PO}}$$^{f}$ 	&f$_{\rm X_{\rm M}}$$^{g}$\\
\hline
\hoii           	&1.42$\pm$0.04      	&0.66$\pm$0.03  &15.73$^{+1.86}_{-1.52}$	&2.67$\pm$0.02      	&2.90$\pm$0.05      	&1463.9/1312		&96.7				&3.3\\
NGC 4395 X-1            &1.73$\pm$0.13      	&0.75$\pm$0.04  &1.32$\pm$0.15      		&3.88$^{+0.08}_{-0.09}$ &5.38$^{+0.46}_{-0.14}$ &{\bf 397.5/366}	&84.4				&15.6\\
NGC 4559 X-1        	&1.26$^{+0.10}_{-0.11}$ &0.25$\pm$0.02  &7.23$^{+1.56}_{-1.54}$ 	&2.25$^{+0.05}_{-0.04}$ &0.24$\pm$0.01      	&{\bf 612.1/597}	&92.8				&7.2\\
\hline
\end{tabular}
\begin{minipage}[t]{6in}
{\sc Notes}: Models are abbreviated to {\sc xspec} syntax: {\sc wa}
and {\sc po} as before, {\sc mekal}--thermal plasma model. $^a$
External absorption column ($10^{21} \atpcm$), $^b$ plasma temperature
(keV), $^c$ {\sc mekal} normalisation (($10^{-19} / (4 \pi
[D(1+z)]^2)) \int n_{\rm e} n_{\rm H} dV$, where D is the distance to
the source (cm), $n_{\rm e}$ is the electron density (cm$^{-3}$),
$n_{\rm H}$ is the hydrogen density (cm$^{-3}$)), $^d$ {\sc po} photon
index, $^e$ {\sc po} normalisation ($10^{-5}~\xspnorm$ at 1 keV), $^f$
fraction of the total flux (0.3--10 keV) in the {\sc po} component,
$^g$ fraction of the total flux (0.3--10 keV) in the {\sc mekal}
component. Statistically acceptable fits are highlighted in bold. (NB
For \hoii we created a dummy response matrix over the 0.3--10 keV
energy range, to determine the fractional flux in each spectral
component (see Sec 4.5)).
\end{minipage}
}
\end{center}
\end{table*}
%=======================================================================

%=======================================================================
\begin{figure}
\begin{center}
\scalebox{0.35}{{\includegraphics[angle=270]{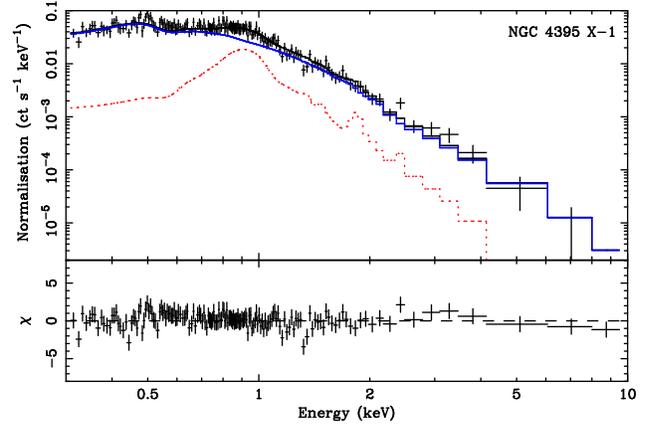}}}
\caption{\label{setpadd}EPIC pn count rate spectra and \delchi
residuals for the {\sc mekal + po} model fit to NGC 4395 X-1.  Also
shown are the individual additive model components: solid
line--{\sc po}, dotted line--{\sc mekal}.}
\end{center}
\end{figure}
%=======================================================================

As shown in the previous sections, statistically acceptable fits for
\hoii and NGC 4395 X-1 cannot be found using the chosen two-component
models.  Interestingly, these are two of the softest sources in the
sample which means that their X-ray spectra are relatively dominated
by $< 1$ keV emission, where it is most sensitive to absorption
characteristics (cf. Fig.~\ref{spec1}).  The analysis of an RGS
spectrum of \hoii by \citet{goad05} revealed that its X-ray emission
is subject to absorption by a medium with a sub-solar oxygen
abundance.  The mis-modelling of absorption, in combination with the
exceptionally high signal to noise EPIC spectrum of Ho II X-1, may be
responsible for the lack of a good fit to this ULX in our analysis.
However, even after correcting for the sub-solar abundance absorption
acting on Ho II X-1, \citet{goad05} still did not find acceptable fits
using the canonical power-law + MCD model.  Instead, a statistically
acceptable solution to the spectrum was found through using a more
physical model, namely the {\sc diskpn + comptt} combination.  We
explore the utility of a similar model in describing our whole dataset
in the next section.

In the case of NGC 4395 X-1, a further inspection of the residuals to
a simple power-law fit does suggest some structure, notably including
a smooth `hump' at $\sim$1 keV, which is unlike the featureless X-ray
spectra of the other ULXs.  Therefore we tried a fit including a {\sc
mekal} component (nominally representing the emission spectrum from
hot, collisionally-ionised gas) in addition to a power-law continuum
to model this spectrum.  The result is shown in Table~\ref{mekalpo}.
This model does indeed provide the best (and only statistically
accceptable) fit to NGC 4395 X-1 with \chisq/dof = 397.5/366.  The
model fitted to the pn count rate spectrum is shown in
Fig.~\ref{setpadd}, as well as the individual additive model
components.  From this figure, one can clearly see how the {\sc mekal}
component effectively models the emission hump at $\sim$ 1 keV.  Using
a fixed solar-abundance absorber, we find the temperature of the {\sc
mekal} component is $\sim$0.75 keV, while the photon index is quite
steep at $\Gamma \sim$ 4.

{\sc mekal} components have been reported in the spectral fits of a
small number of ULXs.  In fact, two other ULXs within this sample have
been described thus in previous analyses, namely \hoii based on joint
\rosat - \asca fits \citep*{miyaji01}, and NGC 4559 X-1 based on one of
two \chandra observations \citep{roberts04}.  We have investigated
whether such {\sc mekal} + power-law models can be applied to the
current data for these sources, and show the results in
Table~\ref{mekalpo}.  This model is again rejected for Ho II X-1,
though it is notable that the quality of the fit is very close to that
of the best empirical description of the source (cool MCD plus
power-law).  In this case the temperature of the {\sc mekal} is again
relatively high at $\sim$ 0.66 keV, but the contribution of this
component is minimal ($\sim$ 3 per cent of the 0.3--10 keV
flux)\footnote{The contribution of the {\sc mekal} is so small that
its parameterisation is relatively insensitive to changes in
metallicity.  For example, setting a low abundance (as found by
\citealt{miyaji01}) leaves the temperature of the {\sc mekal}
unchanged at $\sim$ 0.66 keV.}.  This is both far hotter and far
fainter than the $\sim$ 0.3 keV thermal plasma that composed 20 - 30
per cent of the 0.5--2 keV emission modelled by \citet{miyaji01}.  On
the other hand, the temperature of the best-fitting {\sc mekal} to NGC
4559 X-1 is lower, at $\sim$ 0.25 keV, but it does constitute part of
a statistically-acceptable fit to the data.  This temperature is
similar to the $\sim$ 0.18 keV plasma inferred by \citet{roberts04}.
We note that this latter result just adds to the spectral ambiguity
already seen in the case of this ULX.

%=======================================================================
\begin{table*}
\begin{center}
\small{ \caption{\label{eqpair} {\sc diskpn+eqpair} spectral fits}
\begin{tabular}{lccccccc}
\hline
{\sc wa*wa*(diskpn+eqpair)}&\nh$^a$          &$kT_{\rm max}$$^b$    &A$_{\rm D}$$^c$		&l$_{\rm h}$/l$_{\rm s}$$^d$	&$\tau_{\rm P}$$^e$	&A$_{\rm E}$$^f$              	&\chisq/dof\\
\hline
NGC 55 ULX      &2.12$^{+0.12}_{-0.11}$     &0.24$\pm$0.01          &1.37$^{+0.50}_{-0.36}$     &4.54$^{+1.90}_{-2.44}$ 	&$> 37.2$  		&0.96$^{+3.94}_{-0.24}$    	&{\bf 889.3/827}\\
M33 X-8         &1.03$^{+0.09}_{-0.13}$     &0.08$^{+0.01}_{-0.07}$ &$<4.37 \times 10^{-10}$    &3.22$^{+0.48}_{-0.21}$ 	&18.7$^{+0.6}_{-0.5}$   &1083$^{+21859}_{-78}$ 		&{\bf 1188.0/1112}\\
NGC 1313 X-1    &2.09$^{+0.21}_{-0.39}$     &0.21$^{+0.03}_{-0.02}$ &$<0.44$                	&2.22$^{+2.48}_{-0.46}$ 	&0.2$^{+0.4}_{-0.1}$    &9.99$^{+8.72}_{-2.59}$     	&{\bf 658.7/672}\\
NGC 1313 X-2    &1.44$^{+0.30}_{-0.54}$     &0.26$^{+0.09}_{-0.08}$ &$<0.27$                	&1.83$^{+0.95}_{-0.40}$ 	&9.2$^{+10.7}_{-7.7}$   &0.72$^{+2.80}_{-0.34}$     	&{\bf 254.1/254}\\
NGC 2403 X-1    &2.53$^{+0.23}_{-0.28}$     &0.27$^{+0.05}_{-0.04}$ &0.11$^{+0.21}_{-0.06}$     &4.50$^{+7.39}_{-1.83}$ 	&$> 33.3$               &0.36$^{+1.21}_{-0.31}$     	&{\bf 831.7/848}\\
\hoii		&1.11$^{+0.05}_{-0.07}$     &0.20$^{+0.02}_{-0.01}$ &2.01$^{+0.27}_{-0.19}$     &1.01$\pm0.03$      		&8.5$^{+0.6}_{-0.7}$    &26.7$^{+4.4}_{-8.5}$    	&1414.8/1311\\
M81 X-9     	&1.90$^{+0.14}_{-0.19}$     &0.23$^{+0.01}_{-0.03}$ &$<2.08$                	&3.25$^{+2.09}_{-0.47}$ 	&0.5$^{+0.3}_{-0.1}$    &18.9$^{+16.2}_{-7.0}$   	&{\bf 873.8/861}\\
NGC 3628 X-1    &4.28$^{+0.36}_{-0.42}$     &0.07$^{+0.02}_{-0.01}$ &167$^{+972}_{-112}$   	&26.9$^{+12.6}_{-9.4}$		&$< 7.4$                &47.8$^{+21.1}_{-11.2}$  	&{\bf 424.9/403}\\
NGC 4395 X-1    &1.00$^{+0.12}_{-0.08}$     &0.20$\pm0.01$          &0.15$^{+0.08}_{-0.05}$     &5.97$^{+0.34}_{-3.17}$     	&$> 30.4$   		&0.03$^{+0.03}_{-0.01}$     	&467.0/365\\
NGC 4559 X-1    &1.97$^{+0.15}_{-0.08}$     &0.16$^{+0.01}_{-0.02}$ &2.21$^{+3.57}_{-0.76}$     &2.23$^{+0.51}_{-0.71}$ 	&13.9$^{+3.4}_{-0.6}$   &4.12$^{+1.49}_{-1.45}$     	&{\bf 577.5/596}\\
NGC 4861 ULX    &1.09$^{+0.81}_{-0.42}$     &0.24$\pm0.06$          &0.06$^{+0.22}_{-0.03}$     &5.30$^{+8.65}_{-3.18}$ 	&$> 15.8$  		&0.06$^{+0.55}_{-0.04}$     	&{\bf 78.7/71}\\
NGC 5204 X-1    &0.24$\pm0.08$              &0.29$^{+0.03}_{-0.05}$ &0.09$^{+0.09}_{-0.05}$     &4.32$^{+3.27}_{-1.25}$ 	&26.3$^{+57.0}_{-5.6}$  &0.25$^{+0.60}_{-0.13}$     	&{\bf 505.8/526}\\
M83 ULX         &0.65$^{+0.34}_{-0.33}$     &0.23$^{+0.08}_{-0.07}$ &0.05$^{+0.21}_{-0.04}$     &1.96$^{+7.63}_{-0.85}$ 	&$> 15.8$   		&0.32$^{+3.98}_{-0.19}$     	&{\bf 193.2/204}\\
\hline
\end{tabular}
\begin{minipage}[t]{6.4in}
{\sc Notes}: Models are abbreviated to {\sc xspec} syntax: {\sc wa} as
before, {\sc diskpn}--accretion disc model, {\sc
eqpair}--Comptonisation model, 
$^a$ External absorption column ($10^{21} \atpcm$), 
$^b$ maximum temperature in the accretion disc (keV), 
$^c$ {\sc diskpn} normalisation ($10^{-3} (M^2 \cos(i)) / (D^2 \beta^4)$; 
where M--central mass ($ M_{\odot}$), D--distance to the source (kpc), i--inclination angle of the disc,
$\beta$--colour/effective temperature ratio), 
$^d$ ratio between the compactness of the electrons and the compactness of the seed photon distribution, 
$^e$ optical depth, 
$^f$ {\sc eqpair} normalisation (corresponding to the disc component) \ie $(f_{\rm c} M^2 \cos(i))/(D^2 \beta^4)$, where $f_{\rm c}$ is the covering factor.  
Statistically acceptable fits are highlighted in bold.
\end{minipage}
}
\end{center}
\end{table*}

Finally, if NGC 4395 X-1 really does possess an X-ray line-emitting
component, what is its physical origin?  Several ideas have been put
forward to explain such a component in the spectrum of an ULX.
\citet{miyaji01} suggest the presence of a young supernova remnant
coincident with Ho II X-1, a theme expanded on by \citet{feng05}.  Indeed,
these authors independently confirm the presence of the {\sc mekal}
component in the spectrum of NGC 4395 X-1, and find two further ULXs
with similar characteristics.  Alternately, as \citet{roberts04} find
the {\sc mekal} component in NGC 4559 X-1 to switch on between two
\chandra observations separated by $\sim$ 5 months, they suggest that
the plasma may originate in the collision of a jet or outflow from the
ULX with a denser medium in the close proximity of the system.  A
second alternative is offered by \citet{terashima04}, who suggest that
emission lines in the spectrum of an ULX in M51 may originate from a
photoionised stellar wind, similar to what is seen in some high-mass
X-ray binaries in our own Galaxy.  A final point of interest is that a
broad $\sim$ 1 keV feature, similar to that driving the {\sc mekal}
fit in NGC 4395 X-1, has recently been seen in the \xmmn spectrum of
GRS 1915+105 by \citet{martocchia05} (although see their paper for
caveats).  One intriguing possibility is that this feature could
originate in a disc wind.  Clearly ULXs possessing {\sc mekal}
components in their spectra are an interesting subject in their own
right, and should be the subject of future attention.

%=======================================================================
\subsection{Physical models}

The success of the {\sc comptt} model in describing the spectra of
some ULXs (e.g. \citealt{miller03}, \citealt{goad05}) shows that a
Comptonised spectrum is in principle another viable alternative.  The
{\sc comptt} model however, is not fully self-consistent \eg it allows
spectra to have temperatures higher than expected when Compton cooling
is taken into account.  We therefore attempted to fit a more physical
model to the ULX spectra, namely the {\sc eqpair} model
\citep{coppi99}, which allows thermal and non thermal electron
distributions.  In our modelling, we assume a purely thermal electron
distribution, the temperature of which is computed self-consistently
by balancing heating and cooling (the latter of which is mainly due to
Compton cooling). A key parameter of this model is the ratio $l_{\rm
h}$/$l_{\rm s}$, where $l_{\rm h}$ and $l_{\rm s}$ represent the
compactness of the electrons and the compactness of the seed photon
distribution respectively.  This ratio depends on the geometry of the
source as well as the internal mechanisms, and it provides the
dominant influence on the spectral shape.  The source geometry assumed
is either spherical or a disc-corona (slab) geometry.  Low energy (UV
or X-ray) thermal photons from the accretion disc are assumed to be
emitted uniformly inside the source region for the spherical models
and enter from the base of the corona in the slab geometry.  We take
the seed photons to have an accretion disc spectrum as described with
the {\sc diskpn} model of \citet{gierlinski99}, with the inner edge of
the disc at $6GM/c^2$.  This model assumes a proper general
relativistic potential and has a characteristic temperature $kT_{\rm
bb}$.  To allow for a patchy Compton corona, we add a second {\sc
diskpn} component to the model, with the temperature coupled to that
of the {\sc eqpair} seed photons.

This model ultimately proved the best description of the ULX spectra
in our sample, with statistically acceptable fits to 11/13 datasets,
all with similar or improved goodness of fits compared to the best
empirical modelling, for only one extra degree of freedom.  Only \hoii
and NGC 4395 X-1 were rejected at $> 95$ per cent confidence (see
Table~\ref{eqpair}), although in the case of \hoii it does provide
clearly the best fit of all that were attempted to the data, and is
only marginally rejected ($P_{\rm rej} \approx 97.7$ per cent).  Again
the fit to NGC 4395 X-1 is poor, for reasons explained in the previous
section.

The fits provide a uniformly low measurement of the disc temperature
across the sample, in the range 0.07 $< kT_{\rm max} <$ 0.29.  Taken
in isolation, this might be interpreted as strong support for the
presence of IMBHs in ULXs.  However, many of the fits show a second
remakable characteristic, which is that the optical depth of the
coronae appear to be very high, ranging from $\tau \sim 8$ for \hoii
up to depths well in excess of 30 (NGC 55 ULX, NGC 2403
X-1)\footnote{Indeed, the optical depth for four sources hit the
artificial upper limit of $\tau =$ 100 during spectral fitting,
therefore we quote values as $>$ `lower limit' in these cases.}.
Importantly, this result gives a physical explanation for the
curvature noted to be present in the spectra in previous sections.
This second characteristic appears irreconcilable with the main
assumption behind the IMBH model, which is that they operate as simple
scaled-up BHBs (see also \citealt{goad05}), as such sources typically
do not possess very optically-thick coronae.  This result must
therefore strongly challenge the IMBH interpretation for the nine ULXs
(including Ho II X-1) well-described by this model.  However, three ULXs -
NGC 1313 X-1, M81 X-9 and (possibly) NGC 3628 X-1 - may still possess
optically-thin coronae, and as such remain viable IMBH candidates.

As this spectral model provides at least the same goodness of fit as
the empirical models in all but one case, we used these fits to
determine the observed X-ray flux.  Due to uncertainties in the low energy
spectrum of some ULXs, the spectral fitting was restricted to 0.5--10
keV (and in the case of Ho II X-1 the pn data was restricted to
0.7--10 keV) instead of the 0.3--10 keV range adopted for the other
sources.  Therefore to determine the absorbed X-ray flux of each ULX
over the same energy range, we created a dummy response matrix in {\sc
xspec} over the 0.3--10 keV energy range for those cases with
calibration uncertainties. This dummy response temporarily supersedes
the response matrix used in the spectral fitting of those sources,
allowing us to examine the behaviour of the {\sc diskpn+eqpair} model
over this energy range. The three EPIC cameras are consistent to
within 20 per cent in 12/13 cases, therefore we derived an absorbed
flux (0.3--10 keV) for each of these ULXs based on an average of the
three measurements.  We converted these fluxes to an observed X-ray
luminosity for each source in the 0.3--10 keV band (assuming the
appropriate distance) and tabulate the results in
Table~\ref{sample}\footnote{Although the power-law + {\sc mekal}
spectral model statistically provided the best fit to NGC 4395 X-1,
the measured source flux derived from this model and the {\sc
diskpn+eqpair} model were in close agreement.  For consistency, we
quote the value from the {\sc diskpn+eqpair} model.  However, as the
ULX lay on a hot column in the pn data we simply use the average of
the two MOS flux measurements to estimate its luminosity.}.

%=======================================================================

\section{A comment on the relationship between X-ray luminosity and
inner-disc temperature in ULXs.}
\label{sec_lx_kt}

Perhaps the most visually striking evidence for ULXs containing IMBHs
was presented by \citet*{miller04b} (their Figures 1 \& 2).  These
authors selected a sample of ULXs with published estimates of \lx$ >
10^{40} \ergsec$, that require a soft excess component (at least at
the $3 \sigma$ confidence level) in the low energy part of a two
component X-ray spectrum.  They then compared the inferred disc
temperatures (taken from the soft component) and unabsorbed
luminosities (0.5--10 keV) of their ULXs to those of a number of
Galactic BHBs, and found that these ULXs and stellar mass BHBs occupy
distinct regions of a \lx-- $kT$ diagram. More specifically, the ULXs
are more luminous but have cooler thermal disc components than
standard stellar-mass BHBs, consistent with the ULXs harbouring IMBHs.

%=======================================================================

\begin{figure}
\begin{center}
\scalebox{0.4}{{\includegraphics[angle=270]{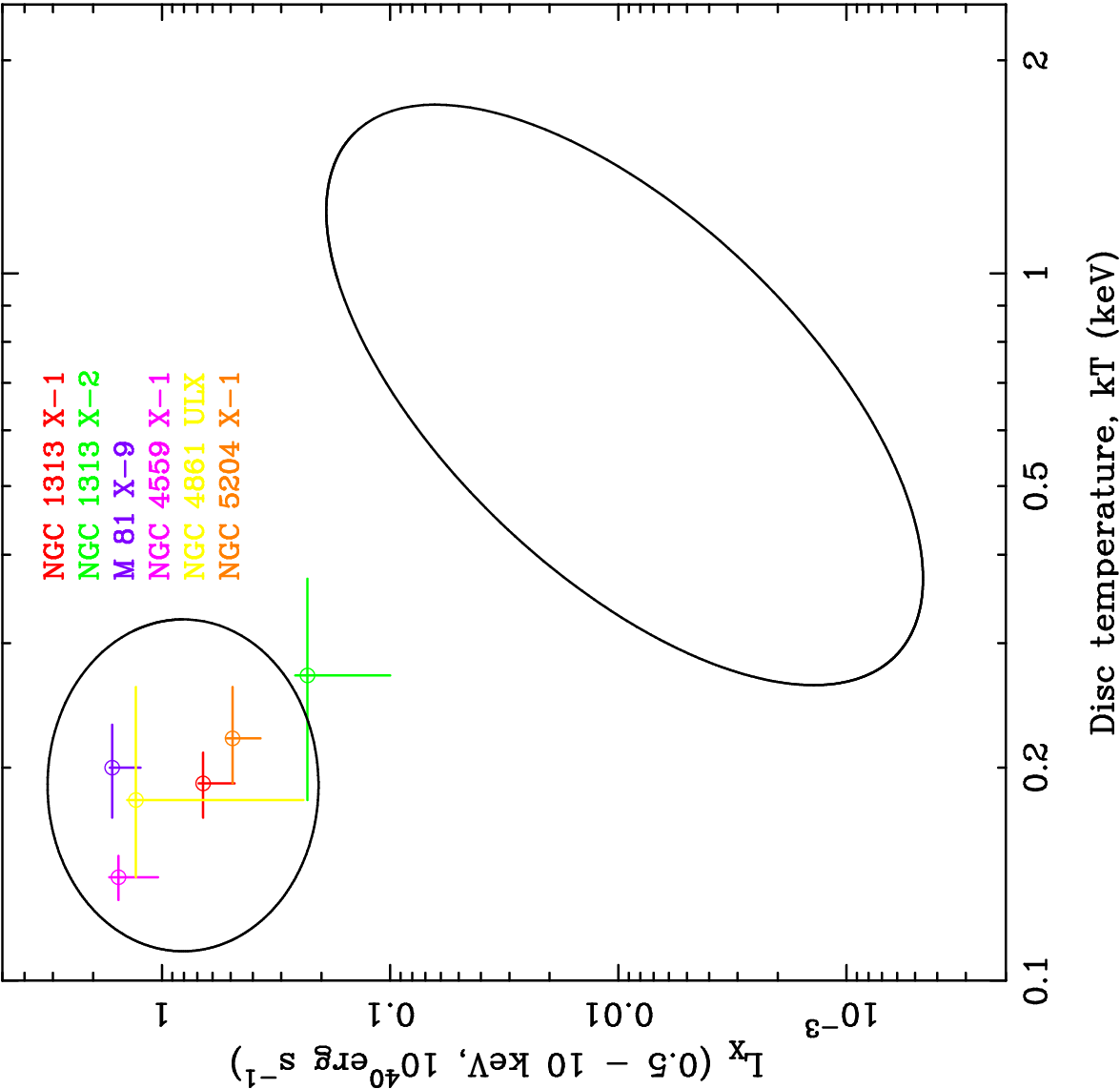}}}\\*[0.61cm]
\scalebox{0.4}{{\includegraphics[angle=270]{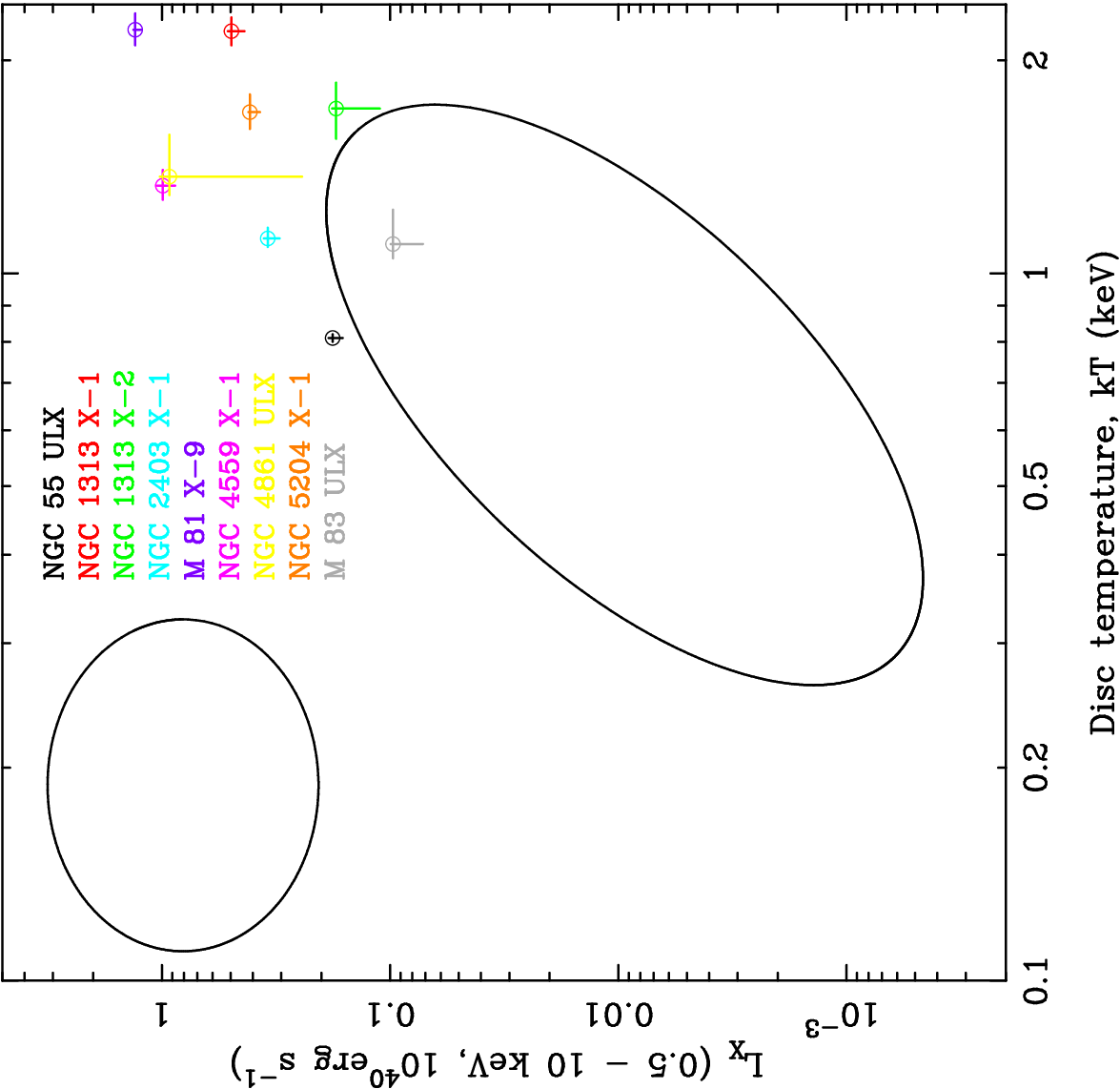}}}
\caption{\label{lx_kt}Unabsorbed X-ray luminosity (0.5--10 keV)
plotted against accretion disc temperatures inferred from the X-ray
spectral fits (as per \citealt{miller04b}). The ellipses represent the
regions occupied by ULXs (small ellipse) and Galactic BHCs (large
ellipse) presented in Figure 2 of \citet{miller04b}.  The values of
the data points displayed in these figures are derived from the IMBH
spectral fits ({\it top panel}) or the dual thermal model fits ({\it
bottom panel}), for sources with statistically acceptable and well
constrained fits for that model. ({\sc Note:} IMBH data points neglect
M83 ULX, which has an unconstrained temperature, and NGC 3628 X-1
which has a lower temperature than the other sources [$\sim$80 eV];
dual thermal data points also neglect NGC 3628 X-1 which has a much
higher temperature than the other sources [$\sim$4 keV]) }
\end{center}
\end{figure}
%=======================================================================

Following on from this work, we have reproduced \lx-- $kT$ diagrams,
based on the inferred unabsorbed 0.5--10 keV X-ray luminosities for
ULXs in this sample (Fig.~\ref{lx_kt}).  However, we do this for two
cases: (i) disc temperatures measured from the IMBH (cool disc + hard
power-law) fits; and (ii) disc temperatures taken from the dual
thermal (cool blackbody + warm disc) fits.  In both cases we only show
sources where the spectral fits are statistically acceptable, and the
fit parameters are well constrained.  For this reason, we excluded the
values from the IMBH fit to M83 ULX, as the disc temperature is
unconstrained in this case.  Additionally, NGC 3628 X-1 does not
appear on either plot as its unusual disc temperatures place it
outside the range displayed in both cases.  In each case the
luminosity was derived from an average of the unabsorbed flux (0.5--10
keV) from each detector. The errors on the disc temperature are 90 per
cent confidence errors, likewise the luminosity errors are based on
the 90 per cent confidence errors in the average measured flux, as
calculated by {\sc xspec}.  Finally, we illustrate the regions of
parameter space occupied by the BHBs and ULXs in \citet{miller04b} by
ellipses, with the ellipse representing IMBH-candidate ULXs at the
upper-left.

The two \lx-- $kT$ diagrams in Fig.~\ref{lx_kt} tell very different
stories.  As would be expected, our IMBH model results reproduce those
of \citet{miller04b}, \ie the ULXs occupy a very different region of
parameter space to the stellar-mass BHBs, consistent with larger black
holes and hence more luminous, cooler accretion discs.  However, the
dual thermal model disc temperatures suggest an alternative
interpretation. In this case, the ULXs appear to be a direct, high
luminosity extension of the BHB class, and follow the \lx$ \propto
T^4$ trend one would expect from standard accretion discs.  Obviously
one cannot conclude which is the correct interpretation on the basis
of these plots.  However, it does demonstrate very clearly that it is
the choice of empirical model used to determine the characteristics of
ULX spectra that governs whether one concludes that some ULXs contain
IMBHs, or whether they are more similar to stellar-mass BHBs.

\section{Discussion}
\label{sec_discussion}

In this work we have analysed the X-ray spectra of a small sample of
13 ULXs, that constituted the highest quality (i.e. most photon-rich)
datasets in the \xmmn data holdings as of December 2004.  We
demonstrate that the superior collecting area of \xmmn over its rivals
leads to better spectral definition, primarily through the rejection
of simple single-component spectral models (power-law, MCD) that have
adequately described previous spectra of many of these ULXs.  However,
we are still photon-limited in many of these spectra, to the extent
that in six cases we are unable to tell whether a combination of a MCD
plus a power-law uniquely fits the data with the MCD component at the
low or the high energy end of the X-ray spectrum.  In each of these
six ambiguous spectra, we accumulated no more than $\sim$ 18000 EPIC
counts (pn and MOS combined).  In each spectrum that we resolved this
simple ambiguity, or excluded both cases as unacceptable fits, we had
$> 20000$ EPIC counts (excluding NGC 4395 X-1, as this had both an
atypical spectrum and an underestimated pn count rate due to its
location on a hot column).  Even then, the introduction of a model
composed of two thermal components adds an extra layer of ambiguity to
many of the datasets.  This suggests that one requires very
photon-rich datasets to make progress in descriptive empirical \xmmn
ULX spectroscopy, equivalent to at least 100-ks of flare-free data for
a 0.2 $\ctsec$ on-axis ULX observed by the EPIC cameras, and spectral
results from poorer quality data should be regarded with caution.

Issues of spectral ambiguity aside, perhaps our most interesting
result is the probable detection of curvature in the high-energy part
of the \xmmn EPIC spectrum in more than half of the sample.  Though
this has previously been detected in individual ULXs
(\citealt{stobbart04}; \citealt{foschini04}), or as minority
populations in ULX samples \citep{feng05}, this is the first
suggestion that the majority of ULXs, across the whole range of ULX
luminosity, might appear thus.  Indeed, we demonstrate that the most
successful empirical modelling of our sample is achieved using a model
in which the high-energy portion of the spectrum is described by the
MCD model, with a soft excess modelled as a cool blackbody emitter.

This obviously adds to the challenges in modelling ULXs as IMBHs.
Indeed, though 8/13 ULXs present acceptable fits to the simple IMBH
model of a cool disc plus a hard power-law continuum, we are again
faced with the problem noted by \citet{roberts05}. Namely we detect a
very dominant power-law ($> 80$ per cent of the observed 0.3--10 keV
flux) that appears too hard for what should be a HS (or perhaps VH)
source, if we are to believe that the inner-disc temperature can
provide an estimate of the BH mass ($1.6 < \Gamma < 2.5$ for the IMBH
candidates in our sample, compared to approximately $2.1 < \Gamma <
4.8$ for the HS, and $\Gamma > 2.5$ for the VH state -
\citealt{mcclintock03}).  Curvature above 2 keV compounds these
problems further, as the IMBH model assumes that whilst the bigger
black hole results in a cooler accretion disc, the corona remains
similar to that observed in Galactic BHBs, i.e. optically-thin and
hence modelled with a power-law.  In fact, there is no reason to
suppose that increasing the mass of the BH should alter the state of
the corona - timing measurements at least appear to scale linearly
with BH mass for given spectral states, implying the properties of the
states themselves are invariant with mass (cf. \citealt{doneg05}).
But the detection of curvature implies that either the state of the
coronae is changing as the BH becomes more massive, or a different
physical process is responsible for the emission above 2 keV.

The dual thermal model uses a MCD to model this curvature, with inner
disc temperatures in the range 1.1--2.2 keV.  As Fig~\ref{lx_kt}
shows, this provides a natural explanation of ULXs as a simple
extension of the behaviour of Galactic BHBs, with more luminous and
slightly hotter discs.  In this case, assuming one can exceed the
Eddington limit, there is no obvious requirement for IMBHs.  However,
even this simple empirical model has limitations; for instance, in
order to see both the inner regions of the accretion disc and an
optically-thick outflow simultaneously, one requires a very specific
geometry and viewing angle.  One may therefore have to consider
further explanations for the origin of the soft excess component; we
note that one interesting suggestion, that has recently been found to
work in some AGN X-ray spectra, is that the soft excess could arise in
relativistically-blurred atomic features found in the reflection
spectrum of a photoionised accretion disc \citep{crummy05}.  A further
concern for this model is that similar spectral shapes have not been
commonly found in Galactic BHBs, with the exception of slight soft
excesses (modelled by the extension of the power-law component below
the MCD) seen in the spectra in LMC X-1 and LMC X-3 \citep{haardt01}.
One explanation could be that soft excesses, below the MCD component
in Galactic BHB spectra, are not seen due to a combination of high
absorption columns and lack of detector sensitivity at low X-ray
energies.  However, the columns to a significant minority of BHBs are
not much in excess of the $\sim 10^{21}$ cm$^{-2}$ columns inferred
for our ULXs \citep{mcclintock03}, and many have now been observed by
\xmmn and \chandra without such soft excesses being reported.  Perhaps
this is simply telling us that this spectral shape is unique to the
very high accretion rates required if ULXs contain stellar-mass BHs.

The model that did best in terms of providing acceptable fits to the
data was also the only physically self-consistent model we used in the
analysis, namely the {\sc diskpn+eqpair} spectral model.  The best
fits provided by this model show, without exception, that the
accretion disc photons seeding the corona are cool.  In fact they are
cool enough, at $< 0.3$ keV, to be consistent with the accretion disc
around an IMBH.  However, this model also provides a physical
explanation for the curvature above 2 keV: unlike in conventional
BHBs, it originates in an optically-thick corona.  This combination of
a cool disc and optically-thick corona has already been observed in Ho
II X-1; \citet{goad05} note the cosmetic similarities between this
spectral model and the model of \citet{zhang00} of a three-layered
atmospheric structure in the accretion discs around BHBs.  More
specifically, their model includes a warm layer ($kT$$\sim$1--1.5 keV,
$\tau$$\sim$10) between the cool optically-thick accretion disc
($kT$$\sim$0.2--0.5 keV) and the hot optically-thin corona
($kT$$\sim$100 keV, $\tau$$\sim$1) of BHBs, which is responsible for
the dominant component below $\sim$10 keV (see also
\citealt{nayakshin97}, \citealt{misra98}).  The warm layer appears
relatively stable and hence unconnected to the hot corona, which can
be highly dynamic and even disappear completely \citep{misra98}.  In
this picture, the cool disc seeds the warm optically thick scattering
medium, and as such may explain both components seen in our modelling.

This model has been successful in describing the X-ray spectrum of GRS
1915+105 \citep{zhang00}, and so \citet{goad05} speculate that this
modelling, combined with an upper limit on the mass of the BH in \hoii
of $\sim 100$-\Msun derived from timing properties, implies that
\hoii behaves in an analogous manner to GRS 1915+105 in its
$\chi$-class.  We speculate, by extension, that many of the ULXs in
our sample that are well fitted by this spectral model could also be
GRS 1915+105 analogues, i.e. stellar-mass BHs accreting at around and
in excess of the Eddington limit.  We note that if individual BHs with
mass up to 80-$M_{\odot}$ can form from stellar processes as suggested
by \cite{belczynski04}, and these larger stellar-mass BHs exist in ULXs,
then the factors by which the Eddington limit are exceeded need not be
large for even the brighter ULXs (factors $\sim 2-3$ only).  One piece
of additional, encouraging evidence in this respect is that, similarly
to \hoii and GRS 1915+105 in its analogous $\chi$-class of behaviour,
the light curves of the ULXs in this sample show little or no
intrinsic variability on timescales of minutes to hours, with the
fractional variability (in excess of counting noise) limited to $\la
10$ per cent in all cases\footnote{Excepting the NGC 55 ULX, which
shows prominent dips in its \xmmn light curve \citep{stobbart04}.}.

Recently a second, very plausible explanation for our physically
self-consistent spectral modelling has come to light.  \citet{donek05}
describe spectral modelling of the Galactic BHB XTE J1550-564 in its
high-luminosity VH state, using a model in which the energetics of the
inner regions of the accretion disc are coupled to a surrounding
corona.  This results in a cooler apparent disc temperature, as the
corona drains energy from the inner disc, and an optically-thick
corona, that are both part of the accretion flow. \citet{donek05} note
that this may provide an explanation for the low disc temperatures
observed in ULXs not reliant upon the presence of an IMBH.  As we also
provide evidence that the corona itself is indeed optically-thick in
such sources, this model must constitute a very serious physical
alternative to IMBHs for the majority of ULXs.  However, even here
there are caveats, for instance \citet{feng05} note that
optically-thick coronae should show deep Fe K absorption edges, unless
the accreting material has a very low metallicity.  We do not detect
such features in our analysis.

Our empirical spectral fitting has posed new challenges for the IMBH
model fits to ULX spectra, and our physical modelling describes
scenarios in which the bulk of ULXs could be stellar-mass BHs
accreting at around the Eddington limit.  This is perhaps not
surprising, as many strands of recent evidence have pointed away from
an IMBH model for most ULXs.  In fact, excepting the somewhat unique
case for M82 X-1 as an IMBH, possibly formed in the dense MGG-11
cluster or captured as the nucleus of an accreted dwarf galaxy
(e.g. \citealt{sm03}; \citealt{pz04}; \citealt{king05};
\citealt{mucciarelli05}), and the cool disc detections (shown to be
somewhat ambiguous in this paper), observational results have tended
to argue in the opposite sense.  For example, the probable breaking of
the Eddington limit seen in some - perhaps most (see
\citealt{jonker04}) - Galactic BHBs, and especially GRS 1915+105,
argues that we cannot exclude stellar-mass BHs from producing ULXs on
this trivial basis \citep{mcclintock03}.  Also, the shape of the
universal X-ray Luminosity Function (XLF) for high-mass X-ray
binaries, derived by \citet{grimm03}, is somewhat puzzling if IMBHs
constitute a significant part of the ULX population.  In particular,
why does the XLF appear to cut-off at $\sim 2 \times 10^{40}
\ergsec$?\footnote{A near identical cut-off is found from a large
sample of ULXs studied by \cite{swartz04}.  Independent support for
this cut-off comes from the empirical $L_{\rm X} -$ star formation
rate relationship of \citet{grimm03}, which can only have its linear
form above $\sim 10 M_{\odot}$ yr$^{-1}$ if the cut-off is real (see
also \citealt{gilfanov04}).}  Surely no such cut-off would be present
if $\sim 1000$-\Msun IMBHs constitute a large fraction of the ULX
population (we are certainly unaware of any other source population
that cuts off at $\sim 0.1$ Eddington rate).  We therefore conclude
that, on current evidence, it is unlikely that accreting IMBHs
constitute a large proportion of the total ULX population.

%=======================================================================

\section{Conclusions}
\label{sec_conclusions} 

We have conducted a detailed examination of the X-ray spectral shapes
in a sample of the highest quality \xmmn EPIC ULX datasets available
to us.  Most notably, more than half of the ULXs show at least
marginal evidence for curvature in their 2--10 keV spectra, which is
somewhat unexpected if they are to be interpreted as the accreting
$\sim 1000$-\Msun IMBHs suggested by modelling the soft spectral
components as accretion discs.  Physical modelling shows that this
curvature is likely to originate in optically-thick coronae, which in
turn leads to interpretations of the ULXs in terms of high
accretion-rate stellar-mass (or slightly larger) BHs operating at
around the Eddington limit.  However, while we conclude that it is
likely that the general ULX population does not have a large
contribution from IMBHs, we obviously cannot rule out the possibility
that some ULXs do possess IMBHs.  Perhaps the best candidate on the
basis of our spectral fitting is M81 X-9, which is well fitted by cool
disc plus power-law/optically-thin corona models, and does not show
explicit curvature in its 2--10 keV spectrum, though even this ULX may
be fitted using a hot ($\sim 2.2$ keV) accretion disc plus soft excess
model.  Clearly, it is difficult to find unique solutions for these
sources even with high quality \xmmn EPIC data.  Ultimately, we may
perhaps have to wait for radial velocity measurements from the optical
counterpart of an ULX, leading to dynamical mass measurements of the
compact accretor, before we have conclusive evidence whether any
individual ULX does harbour an IMBH.
%=======================================================================

\section*{Acknowledgments}

We thank an anonymous referee for their help in improving this paper.
AMS and TPR gratefully acknowledge funding from PPARC.  We thank Mike
Goad for allowing us access to the \hoii data while still proprietary,
and for providing helpful comments on this manuscript.  This work is
based on observations obtained with {\it XMM-Newton} an ESA Science
Mission with instruments and contributions directly funded by ESA
member states and the USA (National Aeronautics and Space
Administration).

\label{lastpage}

%=======================================================================

{}

\end{document}